\documentclass[aoas,preprint]{imsart}

\RequirePackage[OT1]{fontenc}
\RequirePackage{amsthm,amsmath,amssymb}
\RequirePackage{natbib}
\RequirePackage{graphicx}
\RequirePackage{bm, bbm}
\RequirePackage{multirow}

\arxiv{arXiv:0000.0000}

\numberwithin{equation}{section}
\theoremstyle{plain}
\newtheorem{mythm}{Theorem}[section]
\newtheorem{myalg}{Algorithm}[section]

\begin{document}

\begin{frontmatter}
\title{A Multi-Resolution Model for Non-Gaussian Random Fields on a Sphere with Application to Ionospheric Electrostatic Potentials}
\runtitle{Multi-Resolution Model for Non-Gaussian Random Fields}

\begin{aug}
\author{\fnms{Minjie} \snm{Fan}\thanksref{t1,m1}\ead[label=e1]{mjfan@ucdavis.edu}},
\author{\fnms{Debashis} \snm{Paul}\thanksref{t2, m1}\ead[label=e2]{debpaul@ucdavis.edu}}
\author{\fnms{Thomas C. M.} \snm{Lee}\thanksref{t3,m1}\ead[label=e3]{tcmlee@ucdavis.edu}}
\and
\author{\fnms{Tomoko} \snm{Matsuo}\thanksref{t1,m2}
\ead[label=e4]{tomoko.matsuo@colorado.edu}}

\thankstext{t1}{Supported in part by NSF Grants AGS-1025089 and PLR-1443703}
\thankstext{t2}{Supported in part by NSF Grants DMS-1407530, DMS-1713120 and NIH Grant 1R01EB021707 }
\thankstext{t3}{Supported in part by NSF Grants DMS-1512945 and DMS-1513484}
\runauthor{M. Fan et al.}

\affiliation{University of California, Davis\thanksmark{m1} and University of Colorado, Boulder\thanksmark{m2}}

\address{M. Fan\\
D. Paul\\
T. C. M. Lee\\
Department of Statistics\\
University of California, Davis\\
One Shields Avenue\\
Davis, California 95616\\
USA\\
\printead{e1}\\
\phantom{E-mail:\ }\printead*{e2}\\
\phantom{E-mail:\ }\printead*{e3}\\}

\address{T. Matsuo\\
Department of Aerospace Engineering Sciences\\
University of Colorado, Boulder\\
429 UCB\\
Boulder, Colorado 80309\\
USA\\
\printead{e4}}
\end{aug}

\begin{abstract}
Gaussian random fields have been one of the most popular tools for analyzing spatial data. However, many geophysical and environmental processes often display non-Gaussian characteristics. In this paper, we propose a new class of spatial models for non-Gaussian random fields on a sphere based on a multi-resolution analysis.
Using a special wavelet frame, named \textit{spherical needlets}, as building blocks, the proposed model is constructed in the form of a sparse random effects model. The spatial localization of needlets, together with carefully chosen random coefficients, ensure the model to be non-Gaussian and isotropic. The model can also be expanded to include a spatially varying variance profile. The special formulation of the model enables us to develop efficient estimation and prediction procedures, in which an adaptive MCMC algorithm is used. We  investigate the accuracy of parameter estimation of the proposed model, and compare its predictive performance with that of two Gaussian models by extensive numerical experiments. Practical utility of the proposed model is demonstrated through an application of the methodology to a data set of high-latitude ionospheric electrostatic potentials, generated from the LFM-MIX model of the magnetosphere-ionosphere system.
\end{abstract}


\end{frontmatter}

\section{Introduction}\label{sec:intro}

Gaussian random fields (GRF) have provided a very successful modeling framework for analyzing spatial data. Two key ingredients of the success of GRF modeling are: (i) the behavior of the underlying stochastic process is entirely characterized  by the mean and covariance functions, thereby facilitating deep theoretical investigations; and (ii) computations for both estimation and prediction primarily involve matrix algebra. However, there are many geophysical and environmental processes that exhibit a significant degree of non-Gaussianity, such as turbulent fields \citep{Barndorff-79, Berg-16}, meteorological variables including relative vorticity of wind and oceanic currents, wind velocity, air temperature and humidity \citep{Perron-13},  precipitation \citep{De-97, Wallin-15}, and ionospheric electric fields \citep{Cousins-12}.

\subsection{High-latitude ionospheric electric field}
The process that motivates the methodological developments in this paper is the high-latitude ionospheric electric field, originating from solar wind-magnetosphere-ionosphere interactions. Energy and momentum deposition associated with highly variable electric fields leads to global disturbance in the Earth's partially ionized upper atmosphere. This impacts the drag force on low-Earth-orbit satellites and debris, deteriorating our ability of tracking these objects to mitigate potential collisions; and affects radio signal propagation, hindering robust performance of modern technological systems including telecommunication, navigation and positioning. In spite of their scientific importance and societal relevance, current general circulation models of the upper atmosphere are incapable of adequately reproducing these phenomena. One of the obstinate issues originates from systematic underestimation of energy and momentum sources resulting from an inadequate representation of the variability of the high-latitude ionospheric electric fields \citep{Codrescu-95, Matsuo-03}.

The electric fields exhibit considerable variability across a range of scales in both space and time. \citet{Cousins-12} showed in their analysis of the data obtained from 
the \textit{Super Dual Auroral Radar Network (SuperDARN)}, an international network of ground-based high-frequency (HF) radars, that the small-scale spatial and temporal variability of the electric fields displays heavier-tailed behavior than a normal distribution. While a lot of efforts have been made to model the large-scale variability as a GRF through decomposing the process into spherical harmonics (SH) or empirical orthogonal functions (EOF) \citep[see][and references therein] {Cousins-13-Superdarn}, little attempt has been made so far to characterize the non-Gaussian small-scale counterpart.  Although SH and EOF basis functions can well represent large-scale features, they are not effective in capturing small-scale details. 
Moreover, the small-scale variability can form a significant part of the total variability. Through analyzing the SuperDARN data, \citet{Cousins-13} found that together with the mean, the first three EOFs, which are characterized by global spatial scales and long time scales, can only explain approximately $50\%$ of the observed squared electric field. 

Most existing numerical simulations of the upper atmosphere general circulation only account for the large-scale electric fields. Because the amount of Joule heating that results from collisions between neutrals and ions drifting under the effects of the electric and magnetic fields is proportional to the square of the electric field, neglecting the small-scale electric field variability in general circulation models can lead to significant underestimation of energy and momentum inputs into the upper atmosphere. While simple GRF models have been used to parameterize the effects of the large-scale electric field variability on general circulation models \citep{Codrescu-00, Matsuo-08}, there is no viable stochastic parameterization scheme for generating the non-Gaussian small-scale electric fields in a manner consistent with observations. The focus of this paper is therefore to model the small-scale electric field variability for the purpose of incorporating such a model as an adaptive random field generator in general circulation models to account for missing energy and momentum sources. This is a crucial step towards accurately characterizing the variability of the electric fields by random field modeling, and representing the energy budget of solar wind-magnetosphere-ionosphere interactions in numerical models, for facilitating not only further scientific understanding but also practical applications.

\subsection{LFM-MIX model output}
We consider a simulated data set of high-latitude ionospheric electrostatic potentials generated from the LFM-MIX model, a state-of-the-art coupled magnetosphere-ionosphere model \citep{Lyon-04} capable of running at multiple resolutions \citep{Wiltberger-16}. Note that the aforementioned electric fields are electrostatic fields given as the negative gradient of the electrostatic potentials. The domain of the LFM-MIX model relevant to magnetosphere-ionosphere coupling is the high-latitude regions (i.e., latitude $\geq 45^\circ$ or $\leq -45^\circ$) of the Northern and Southern Hemispheres. LFM-MIX global simulation results were compared to real observations and empirical models of high-latitude ionospheric electrodynamics, and good agreement was reported \citep{Zhang-11, Wiltberger-16, Kleiber-16}. \citet{kleiber-13} and \citet{heaton-15} constructed spatio-temporal statistical emulators of the LFM-MIX model, and using the emulators, they estimated the input parameters of the LFM-MIX model  through matching the model output with real observations.

\begin{figure}[htbp] %  figure placement: here, top, bottom, or page
   \centering
   \includegraphics[width=3.5in]{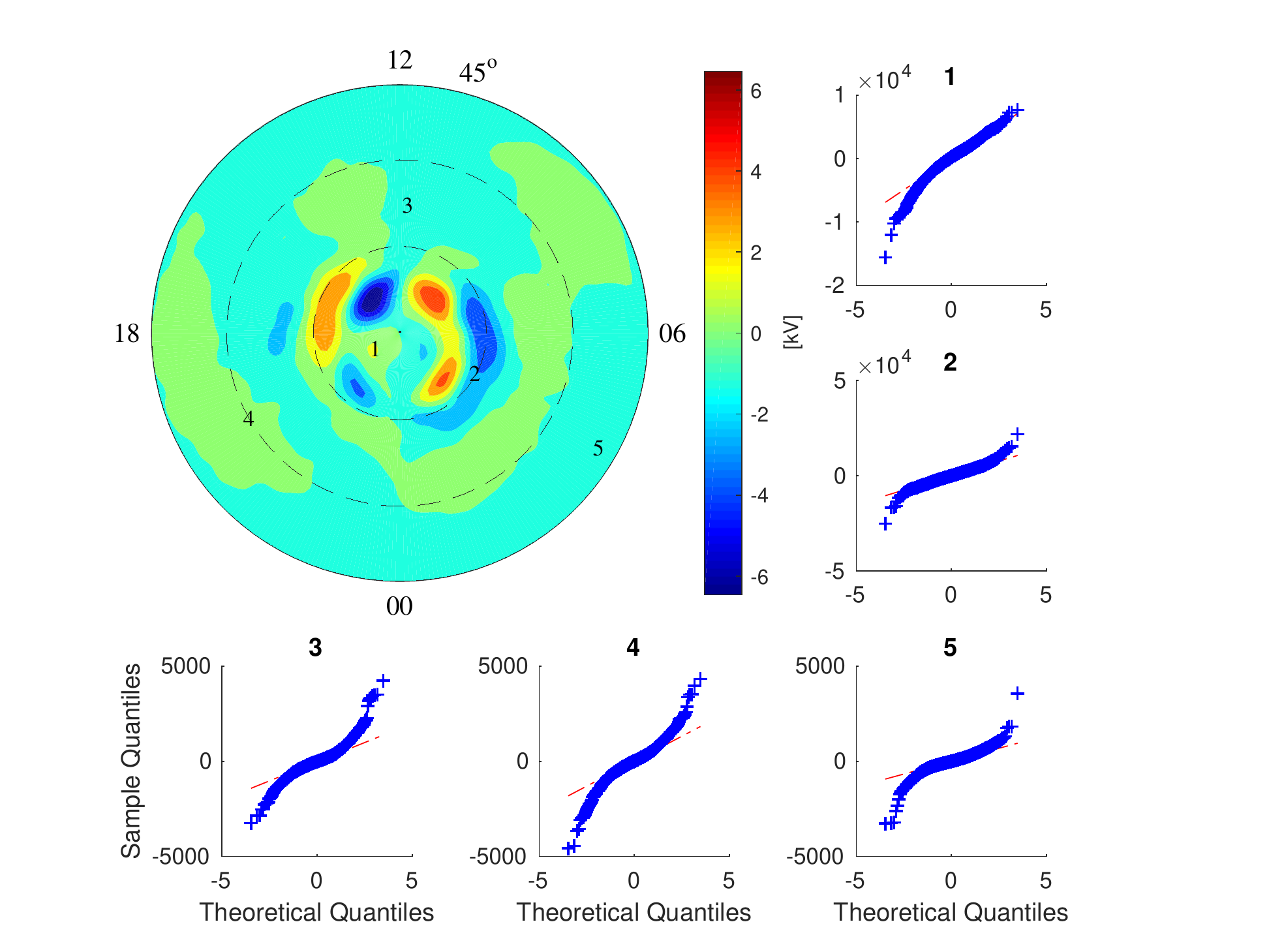} 
   \caption{Small-scale component of the high-latitude ionospheric electrostatic potentials in the Northern Hemisphere generated from the LFM-MIX model at the Quad resolution. \textbf{Top left panel:} the small-scale component at the first time point; \textbf{Other panels:} the non-Gaussianity of the small-scale component is illustrated by the Q-Q plots of the replications over time at different locations.}
   \label{fig:real_nonGau}
\end{figure}

Global simulations using the LFM-MIX model especially at higher resolutions require a considerable amount of high-performance computing resources, which by itself provides another motivation for the stochastic parameterization of the small-scale electric field variability for further numerical investigations using upper atmosphere general circulation models. Herein we focus on the Quad resolution output of the LFM-MIX model in the high-latitude region of the Northern Hemisphere since the model output in the Southern Hemisphere can be analyzed similarly. Through an exploratory analysis of the Quad resolution model output, we found that the large-scale features (i.e., the mean and the first four EOFs) of the electrostatic potentials in the Northern Hemisphere show patterns similar to those derived from real observations in \citet{Matsuo-02, Cousins-13} (see Section S.2.1 of the supplementary material \citet{Fan-17-supp}). Figure \ref{fig:real_nonGau} shows the small-scale component of the high-latitude ionospheric electrostatic potentials in the Northern Hemisphere after subtracting the estimated large-scale component (see Section \ref{sec:data_preproc}). The non-Gaussianity of the small-scale component is illustrated by the Q-Q plots of the replications over time at different locations. Thus, the small-scale component of the high-latitude ionospheric electrostatic potentials is naturally a non-Gaussian random field on a sphere. 
 
\subsection{Challenges and contributions}
One popular approach for dealing with non-Gaussian processes is to transform the data by some non-linear function such that they can be modeled by a GRF. See \citet{Cressie-93, De-97, Xu-16} for examples. The success of this approach relies heavily on finding a suitable transformation for the data. As the covariance structure of the latent GRF becomes more complicated, the transformation could make the derived non-Gaussian process more difficult to interpret.  Among various approaches to modeling non-Gaussian processes directly without any transformation, \citet{Palacios-06} proposed a class of non-Gaussian spatial models based on scale mixing of a Gaussian process; \citet{Roislien-06} developed a $t$-distributed random field model with heavy-tailed marginal probability density functions, which is a generalization of the familiar multivariate $t$-distribution; and \citet{Wallin-15} derived non-Gaussian geostatistical models with a Mat\'ern covariance structure from stochastic partial differential equations driven by non-Gaussian noise.
%Due to the difficulty in both theory and computation, the modeling of non-GRFs is one of the frontier challenges in spatial statistics.

Modeling random fields on a sphere instead of a Euclidean space poses an additional technical challenge. Processes on a sphere are usually constructed by restricting processes on $\mathbb{R}^3$ to the sphere, but this may cause physically unrealistic distortions, especially for the covariance structure at long distances \citep{Gneiting-13}.
There have been efforts to develop valid covariance functions on a sphere directly (see \citet{Jun-08, Guinness-16} for examples), but having a covariance function only does not suffice to fully define a non-Gaussian process. \citet{Stein-07} and  \citet{Cressie-08}, among others, used a fixed rank approach in which GRFs on a sphere are approximated by a linear combination of basis functions with normally distributed coefficients. The stochastic properties of the resulting process are determined by the distribution of the coefficients and the choice of the basis functions.

In this paper, we propose a new class of multi-resolution spatial models for non-Gaussian random fields on a sphere, which are constructed to mimic the characteristics of the aforementioned  high-latitude ionospheric electrostatic potentials. 
The construction of the model starts with a sparse random effects model, represented in terms of a special multi-resolution wavelet frame named spherical needlets. The spatial localization of needlets, together with carefully chosen random coefficients in this representation, ensure that the resulting process is non-Gaussian and isotropic. The model is then expanded to include a spatially varying variance profile. Motivated by the specific application, we consider a special case where the variance depends on the latitude only. Analytical properties of needlets and the special formulation of the model enable us to develop efficient estimation and prediction procedures. 

We apply the proposed model to the simulated high-latitude ionospheric electrostatic potentials generated from the LFM-MIX model. The results reveal that the amount of Joule heating is significantly increased by taking into account and modeling the small-scale variability of the electrostatic potentials using the proposed model. This indicates the necessity of including the small-scale variability in the calculation of the Joule heating. In comparison, modeling the small-scale variability using the popular Gaussian Mat\'ern model fails to produce significantly more energy than the large-scale component only. In current general circulation models, the Joule heating rate is underestimated and hence often multiplied by an arbitrary factor such that the latitudinal distribution of the modeled upper atmosphere temperature matches well with observed climatology \citep{Matsuo-08}. However, such an ad-hoc treatment of the energy source has a number of shortcomings. In particular, since the electric fields also drive winds through the ion-drag force, the energy and momentum resulting from the high-latitude electric fields cannot be arbitrarily altered and need to be consistently treated. We also notice that the proposed model has a higher chance of producing extremely high energy than its Gaussian version due to the non-Gaussianity of the model.
In fact, the upper atmosphere responds to extreme energy and momentum sources differently from moderately elevated ones. Thus, the proposed model's ability to produce extreme energy and momentum sources is promising to improve the accuracy of numerically modeling the magnetosphere-ionosphere coupling.
 
The remainder of the paper is organized as follows. In Section \ref{sec:spatial_model}, we first give a brief introduction to spherical needlets, and using needlets as building blocks, we then construct a sparse random effects model for non-Gaussian random fields on the unit sphere. The model is also expanded to include a spatially varying variance profile. In Section \ref{sec:comp_detail}, we describe the computational details of model fitting, prediction and unconditional simulation. Their good performance is demonstrated in Section \ref{sec:num_exp} by extensive numerical experiments. In Section \ref{sec:app}, we apply the proposed model to the high-latitude ionospheric electrostatic potentials. Some relevant issues and future directions are discussed in Section \ref{sec:discuss}.

\section{Model construction for non-Gaussian random fields\label{sec:spatial_model}}

In this section, we construct a class of spatial models for non-Gaussian random fields on the unit sphere through a sparse random effects model that uses 
a multi-resolution representation in the form of a spherical needlet frame.
  
\subsection{Spherical needlets}
We first give a brief introduction to spherical needlets and some of their important properties. More details can be found in \citet{Narcowich-etal06, Marinucci2011, Fan-15}. 

Let $\mathbf{s}$ denote a point on the unit sphere $\mathbb{S}^2=\{\mathbf{x}\in \mathbb{R}^3: \lVert \mathbf{x} \rVert=1 \}$, and $(\theta, \phi)$ represent the same point in spherical coordinates, where $\theta$ and $\phi$ are the co-latitude and longitude, respectively. Complex-valued spherical harmonics (SH), denoted by $\{ Y_{lm}(\theta, \phi), l=0,1,\cdots,m=-l,\cdots, l\}$, are 
the spherical analogue of the Fourier basis on the unit circle. They form an orthonormal basis for the Hilbert space $L^2(\mathbb{S}^2)$, the space of square integrable functions on the unit sphere. The index $l$ determines the frequency level of SH functions.

Needlets are constructed in terms of SH functions based on two key ideas,
namely, (a) a discretization of the unit sphere; and (b) a Littlewood-Paley decomposition. The discretization of $\mathbb{S}^2$ is achieved by an exact \textit{quadrature formula}: for every $l\in \mathbb{N}$, there exist a finite subset 
$\mathcal{X}_l=\{ \bm{\zeta}_{lk} \}_{k=1}^{n_l}\subset \mathbb{S}^2$ (\textit{quadrature points}) and positive weights $\{ \lambda_{lk}\}_{k=1}^{n_l} $ (\textit{quadrature weights}) such that, for any polynomial $f$ of degree at most $l$,
\begin{equation}\label{quadature}
\int_{\mathbb{S}^2} f(\mathbf{s})d\mathbf{s} = \sum \limits_{k=1}^{n_l} \lambda_{lk} f(\bm{\zeta}_{lk}).
\end{equation}
The Littlewood-Paley decomposition is defined through a function $b$ on
$\mathbb{R}^+$ satisfying:
(i) $b(\cdot) > 0$ on $(B^{-1},B)$ for some $B > 1$, and equal to zero on $(B^{-1},B)^c$; (ii)
$\sum_{j=0}^\infty b^2(y/B^j) = 1$ for all $y \geq 1$; and (iii) $b(\cdot) \in C^M(\mathbb{R}^+)$ for some $M \in \mathbb{N} \cup \{\infty\}$.
Based on these specifications, a class of spherical needlets is defined as follows. For $\mathbf{s} \in \mathbb{S}^2$, 
\begin{eqnarray}\label{needlet}
\psi_{jk}(\mathbf{s}) & = & \sqrt{\lambda_{jk}} \sum \limits_{l=\lceil B^{j-1} \rceil}^{\lfloor B^{j+1} \rfloor} b\left( \frac{l}{B^j} \right) \sum \limits_{m=-l}^{l} Y_{lm}(\bm{\zeta}_{jk})\overline{Y}_{lm}(\mathbf{s}) \nonumber \\
& = & \sqrt{\lambda_{jk}} \sum \limits_{l=\lceil B^{j-1} \rceil}^{\lfloor B^{j+1} \rfloor} b\left( \frac{l}{B^j} \right) \left( \frac{2l+1}{4\pi} \right) P_l(\langle \bm{\zeta}_{jk}, \mathbf{s} \rangle),
\end{eqnarray}
where $j\in \mathbb{N}\cup\{ 0\}$ encodes the \textit{scale} or \textit{frequency} of needlets, (with a slight abuse of 
notation) $\{\bm{\zeta}_{jk}\}_{k=1}^{p_j}$ are a set of quadrature points, and $\{\lambda_{jk}\}_{k=1}^{p_j}$ are the 
corresponding quadrature weights, where $p_j = n_{C_j}$ with $C_j = 2 \lfloor B^{j+1}\rfloor$
and $n_l$ is as in (\ref{quadature}). The point $\bm{\zeta}_{jk}$ determines the \textit{location} of the 
needlet $\psi_{jk}$, for each $k=1,\ldots,p_j$. Besides, $\langle \cdot, \cdot \rangle$ denotes the dot product on $\mathbb{R}^3$, and $P_l$ represents the $l$-th Legendre polynomial. Note that the last equality holds due to the \textit{Addition Theorem for SH functions} \citep[Theorem 2.9]{Atkinson-12}, and it also shows that needlets are real-valued functions.

Needlets possess several attractive properties. They are localized in both the spatial and frequency domains, with
quasi-exponentially increasing concentration around the quadrature point $\bm{\zeta}_{jk}$ as the frequency level $j$ increases.
Together with $Y_{00}$, a constant function on the unit sphere, the collection of needlets also form a
\textit{Parseval tight frame}, i.e., for any  $f \in L^2(\mathbb{S}^2)$,
$$\int_{\mathbb{S}^2} |f(\mathbf{s})|^2 d\mathbf{s} = |\langle f,Y_{00}\rangle_{L^2}|^2 + \sum_{j=0}^\infty \sum_{k=1}^{p_j} |\langle f,\psi_{jk}\rangle_{L^2}|^2,$$
where $\langle \cdot,\cdot \rangle_{L^2}$ denotes the inner product on $L^2(\mathbb{S}^2)$, and
$$\langle f,\psi_{jk}\rangle_{L^2} = \int_{\mathbb{S}^2} f(\mathbf{s}) \psi_{jk}(\mathbf{s})d\mathbf{s}$$ is called the
\textit{needlet coefficient} of $f$ corresponding to the index pair $(j, k)$, denoted by $\beta_{jk}$. The tight frame property, together with the localization in both the spatial and frequency domains, ensure that needlets can be used to perform a \textit{multi-resolution analysis} of functions in $L^2(\mathbb{S}^2)$. 

In our construction, we use the symmetric spherical $t$-designs on $\mathbb{S}^2$ \citep{Womersley-15} as the quadrature points. The number of quadrature points $n_l=l^2/2+l/2+\mathcal{O}(1)$, and the quadrature weights are all equal to $4\pi/n_l$. Moreover, the function $b$ is chosen based on the second specification in \citet[Chapter 10.2.2]{Marinucci2011} with $b(\cdot) \in C^{\infty}$ (see Section S.1.1 of the supplementary material \citet{Fan-17-supp} for details), and $B$ is specified as $2$ according to \citet{Narcowich-etal06}.

\subsection{Sparse random effects model}\label{sec:spa_rand_eff}

In this subsection, we construct a sparse random effects model for scalar random fields on the unit sphere using needlets as building blocks. Let $\{X(\textbf{s}): \textbf{s}\in \mathbb{S}^2\}$ be a zero-mean scalar random field on the unit sphere, which can be represented in terms of needlets, i.e., 
\begin{equation}\label{w_comp}
X(\textbf{s})=\sum \limits_{j=J_0}^{J} \sum \limits_{k=1}^{p_j} c_{jk}\psi_{jk}(\textbf{s}),
\end{equation}
where $J\geq J_0 \geq 0$ determine the frequency range of the information contained in the process. In general, the representation is not unique due to the fact that $\{\psi_{jk} \}_{j,k}$ form a tight frame, which is an overcomplete system. Thus, we need to impose certain probabilistic assumptions on the random coefficients $c_{jk}$'s such that meaningful inference on model parameters is enabled with pooling of information. Another consideration is to make the process $X$ non-Gaussian. This can be achieved through assuming that 
$c_{jk}$'s have a non-Gaussian joint probability distribution. In order to incorporate both scale-dependent variations and non-Gaussianity, we make the following key structural assumptions:

\begin{itemize}
\item
For each index pair $(j,k)$, there are parameters $\nu_{jk} > 2$ and $\sigma_{jk} > 0$ such that $c_{jk}
\sim \sigma_{jk} t(\nu_{jk})$, where $t(\nu)$ denotes the $t$-distribution with $\nu$
degrees of freedom. Besides, $c_{jk}$'s are independent.
\end{itemize}
The independence assumption of the coefficients $c_{jk}$'s is motivated by the fact 
that, for a two-weakly isotropic process with mild regularity conditions,
the correlation among pairs of needlet coefficients decay rapidly as the corresponding quadrature points and frequency levels become more separated \citep{Baldi-09, Marinucci2011}.

The $t$-distribution assumption and the spatial localization of needlets ensure non-Gaussian stochastic properties of the process $X$ because for any given location $\mathbf{s}$, $X(\mathbf{s})$ is approximately a weighted sum of a small number of $c_{jk}$'s, which are independent and non-Gaussian.
The same does not hold if needlets are replaced by basis functions with global support, such as SH functions, since in that case, $X(\mathbf{s})$ becomes a weighted sum of a large number of $c_{jk}$'s and the central limit theorem kicks in, resulting in approximately Gaussian behavior. This specification also has a distinct computational advantage, which will be discussed in Section \ref{sec:MCMC}. 

We have the following theorem characterizing the covariance structure of the process $X$ in a special case.
\begin{mythm}\label{cov_fun}
If we consider the special case where $\sigma_{jk} = \sigma_j$ and $\nu_{jk} = \nu_j$ with $\nu_j>2$ for all $k=1,\ldots,p_j$,
the resulting process $X$ is two-weakly isotropic with an oscillating covariance function
\begin{equation}\label{osc_cov_fun}
C(\mathbf{s}, \mathbf{t}) = \mbox{Cov} \left(X(\mathbf{s}), X(\mathbf{t})\right)=\sum \limits_{j=J_0}^J  \frac{\nu_j\sigma_j^2}{\nu_j-2}\sum_l b^2 \left( \frac{l}{B^j} \right)\left(\frac{2l+1}{4\pi}\right)
P_l(\langle \mathbf{s}, \mathbf{t} \rangle),
\end{equation}
where $\mathbf{s}, \mathbf{t} \in \mathbb{S}^2$.
The covariance function decays quasi-exponentially with respect to the great-circle distance between the two locations $\mathbf{s}$ and $\mathbf{t}$
\begin{equation}\label{decay}
\lvert C(\mathbf{s}, \mathbf{t}) \rvert \leq \sum \limits_{j=J_0}^J  \frac{\nu_j\sigma_j^2}{\nu_j-2} \frac{c_M B^{2j}}{[1+B^j \arccos (\langle \mathbf{s}, \mathbf{t} \rangle)]^M},
\end{equation}
where $c_M$ is some constant depending on $M$. Obviously, when $b(\cdot) \in C^{\infty}$, (\ref{decay}) holds for all $M\in \mathbb{N}$.
\end{mythm}
The proof is deferred to Appendix \ref{sec: proof_prop1}. The isotropy of the process is mainly due to the assumption that $\sigma_{jk}$ and $\nu_{jk}$ do not vary with respect to $k$. Extensions to anisotropic models are possible by allowing $\sigma_{jk}$ and $\nu_{jk}$ to vary spatially, e.g., $\sigma_{jk}=f(\bm{\zeta}_{jk})\sigma_j$ and $\nu_{jk}=g(\bm{\zeta}_{jk})\nu_j$ for certain functions $f$ and $g$. For simplicity, in the rest of the paper, we focus on the special case where $\sigma_{jk}=\sigma_j$ and 
$\nu_{jk}=\nu_j=\nu$.

There are some connections of the derived covariance function (\ref{osc_cov_fun}) with existing literature. \citet{Lindgren-11} discussed an approach for constructing scalar GRFs on $\mathbb{R}^d$ and $\mathbb{S}^2$ with oscillating covariance 
functions through stochastic partial differential equations. These covariance functions have an oscillating structure similar to our proposal, but they do not necessarily decay 
quasi-exponentially with respect to the great-circle distance. Moreover, there is no closed-form expression for the covariance functions when they are defined on $\mathbb{S}^2$. \citet{Schoenberg-42} showed that $\{C(\theta): 0\leq \theta \leq \pi\}$ is a valid isotropic continuous covariance function on $\mathbb{S}^2$ if and only if it has the following representation in terms of Legendre polynomials
$$
C(\theta)=\sum \limits_{l=0}^\infty a_lP_l(\cos \theta),
$$
where $\theta$ is the great-circle distance between two locations, $a_l\geq 0$ for all $l\geq 0$ and  $\sum_{l=0}^\infty a_l<\infty$. 
This general form was used by \citet{Terdik-15} and \citet{Guinness-16} through specifying specific $a_l$'s to construct isotropic covariance functions on $\mathbb{S}^2$.

We now consider an observation model based on the above formulation. Suppose that we have observations on $X$ at $n$ different locations, $\mathbf{s}_1, \cdots, \mathbf{s}_n$, and these observations, denoted by $Z(\mathbf{s}_i)$, are corrupted by observational errors. Then
\begin{equation}\label{obs_model}
Z(\mathbf{s}_i)=X(\mathbf{s}_i)+e_i, \quad i=1,\cdots, n,
\end{equation}
where $e_i$'s are the observational errors modeled as i.i.d. $\mathcal{N}(0, \tau^2)$. In practice, we can specify the parameter $J$ as some value $J_{\rm max}$ based on data resolution. The parameter $\sigma_j$ usually decays as the frequency level $j$ increases. For example, we may assume that $\sigma_j=f(j)\sigma$, where $f(j)=B^{-\alpha j/2}$ with $\alpha>2$ (see Section S.1.2 of the supplementary material \citet{Fan-17-supp} for justification). The hyperparameter $\alpha$ controls the decay rate of the magnitude of the coefficients $c_{jk}$'s from low to high frequency levels. 
In general,  $\sigma_j$ can be estimated without imposing any particular structure.

\begin{figure}[htbp] %  figure placement: here, top, bottom, or page
   \centering
   \includegraphics[width=3.5in]{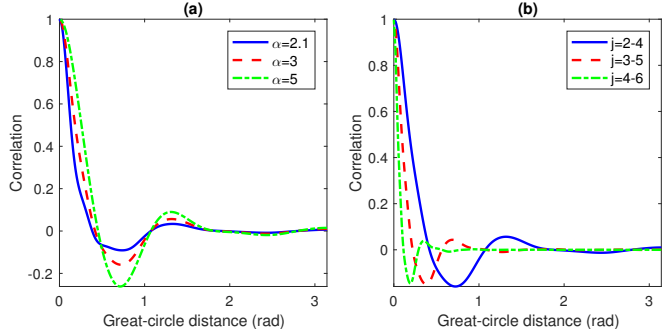} 
   \caption{\textbf{(a)} Correlation functions for various $\alpha$ with $j$ ranging from $2$ to $4$; \textbf{(b)} Correlation functions for various ranges of $j$ with $\alpha=3$.}
   \label{fig:corr_fun}
\end{figure}

In Figure \ref{fig:corr_fun}, we give a few examples of the correlation function of the process $X$ with $\sigma_j=2^{-\alpha j/2}$ for various $\alpha$ and ranges of $j$. The curve oscillates more dramatically as $\alpha$ increases, and decays more rapidly as $J_0$ becomes larger. The parameter $\nu$ has no impact on the correlation structure, but it measures the degree of non-Gaussianity of the process. In all the cases, the curve decays quasi-exponentially fast as we have shown in Theorem \ref{cov_fun}.

To demonstrate the effect of the non-Gaussianity assumption for the coefficients $c_{jk}$'s, we simulate the process $X$ when $c_{jk}$ is distributed as $\sigma_jt(\nu)$ and normally distributed with the same variance, respectively. We specify $\sigma_j=2^{-\alpha j/2}$ with $\alpha=3$, $J_0=2$, $J=4$ and $\nu=2.5$. Figure 9 in Section S.3 of the supplementary material \citep{Fan-17-supp} shows the projection of the simulations onto frequency levels $j=3, 4$, i.e., $X^{(j)}:=\sum_{k=1}^{p_j}c_{jk}\psi_{jk}$. We can see that the non-Gaussian coefficients yield more positive and negative extreme values than the Gaussian ones. The Q-Q plots of $10,000$ i.i.d. simulations of the process $X$ at a specific location, displayed in Figure 10 in Section S.3 of the supplementary material \citep{Fan-17-supp}, confirm its non-Gaussianity when $c_{jk}$'s are $t$-distributed. Checking the non-Gaussianity at one location is sufficient due to the isotropy of the process.

\subsection{Axially symmetric model with a variance profile}

Spatial data on a global scale usually exhibit nonstationary behavior. 
In this subsection, we therefore expand the sparse random effects model to include a spatially varying variance profile. Let $\{g(\mathbf{s}), \mathbf{s}\in \mathbb{S}^2\}$ be a function that characterizes the spatially varying variance structure.
The function $g$ is modeled through a log-linear basis representation
\begin{equation}
g(\textbf{s})= \exp \left(\mathbf{b}^{\rm T}(\textbf{s})\bm{\eta} \right),
\end{equation}
where $\mathbf{b}(\textbf{s})=(1, b_1(\textbf{s}), \cdots, b_r(\textbf{s}))^{\rm T}$ is a vector of basis functions (including the intercept) evaluated at location $\mathbf{s}$, and $\bm{\eta}=(\eta_0, \bm{\eta}_{-0}^{\rm T})^{\rm T}$ is the corresponding coefficient vector. Through an exploratory analysis of the LFM-MIX model output, we notice that the variance of the high-latitude ionospheric electrostatic potentials shows a strong dependence on the latitude, while only a moderate dependence on the longitude. Motivated by this, for simplicity and interpretability, we ignore the longitudinal dependence, and assume that $g(\textbf{s})$ is a function of the co-latitude $\theta$ only, i.e., 
\begin{equation}\label{gs}
g(\textbf{s})\equiv g(\theta)=\exp \left(\mathbf{b}^{\rm T}(\theta)\bm{\eta} \right).
\end{equation}
Moreover, the variance is assumed to vary smoothly over $\mathbb{S}^2$, and therefore the basis functions are specified as cubic B-splines due to their numerical stability, with the first B-spline replaced by the intercept. Then, we have the following non-Gaussian model with a variance profile
\begin{equation}\label{AXING}
X(\textbf{s}) = g(\theta)\sum \limits_{j=J_0}^{J}\sum\limits_{k=1}^{p_j}c_{jk}\psi_{jk}(\textbf{s}).
\end{equation}
Since the function $g$ depends on the latitude only, this model is axially symmetric \citep{Jones-63}. 
We refer to the model given by (\ref{gs}) and (\ref{AXING}) as the \textit{axially symmetric non-Gaussian needlet model},
abbreviated as \textit{AXING-need}. As a generalization of the model, 
one may use known functions of spatial covariates instead of $\mathbf{b}(\theta)$ in the log-linear basis representation (\ref{gs}) if they are available. 

We now consider the simple observation model (\ref{obs_model}) with $X$ defined by (\ref{AXING}). Denote the vector of observations by $\mathbf{Z}=(Z_1, \cdots, Z_n)^{\rm T}$, where $Z_i=Z(\mathbf{s}_i)$, and the vector of observational errors by $\mathbf{e}=(e_1, \cdots, e_n)^{\rm T}$. The AXING-need model can be written in matrix-vector form 
\begin{equation}\label{AXING_matrix}
\textbf{Z}=\textbf{G} \textbf{A} \textbf{c}+\textbf{e},
\end{equation}
where $\textbf{G}:=\mbox{diag} \left\{ g(\textbf{s}_1), \cdots, g(\textbf{s}_n) \right\}$, $\mathbf{A}$ is the $n$-by-$\sum_j p_j$ design matrix with the $i$-th row corresponding to the value of needlets at $\mathbf{s}_i$, and columns ordered according to the index pair $(j, k)$, and $\mathbf{c}$ is the coefficient vector consisting of $c_{jk}$'s.

\section{Model fitting and prediction}\label{sec:comp_detail}

In this section, we describe the computational details of model fitting, prediction and unconditional simulation. The latter two are used to test the performance of the proposed AXING-need model, and compare it with Gaussian models. Computing maximum likelihood estimates (MLE) in the AXING-need model is intractable due to the absence of a closed form of the likelihood for the parameters. Evaluation of the likelihood requires high-dimensional integrals over the distribution of the random coefficients $c_{jk}$'s. Thus, we adopt a Bayesian approach, and implement a Markov Chain Monte Carlo (MCMC) algorithm that samples from the posterior distribution of the parameters. Before going into the details, we impose the following structural restrictions for the ease and stability of the computation:
(i) the parameter $\nu$ is fixed and known \citep{Wolfe-04}; and (ii) the parameters $\eta_0$ and $\sigma_{J_0}$ are non-identifiable. We choose to fix $\eta_0$ at 0 and treat $\sigma_{J_0}$ as a free parameter, so that the latter can be sampled in a Gibbs step. According to our numerical experiments, this is more efficient than sampling $\eta_0$. We denote the vector of the remaining parameters by $\bm{\theta}=(\sigma_{J_0}^2, \cdots, \sigma_{J}^2, \tau^2, \bm{\eta}_{-0}^{\rm T})$.

In the following derivation of the algorithm, generic notations are used: $[U]$ denotes the distribution (or density) of a random variable $U$; and $[U|V]$ denotes the conditional distribution (or density) of $U$ given $V$.

\subsection{Prior specification}

We assume that the parameters are \emph{a priori} independent, i.e., $[\bm{\theta}]=[\sigma_{J_0}^2]\cdots[\sigma_J^2][\tau^2][\bm{\eta}_{-0}]$.
The prior distributions of $\sigma_j^2$ and $\tau^2$ are specified as the non-informative Jeffreys' priors $[\sigma_j^2]=1/\sigma_j^2$ and $[\tau^2]=1/\tau^2$.
The prior distribution of $\bm{\eta}_{-0}$ is assumed to be $\mathcal{N}(\bm{0}, \tau^2_{\bm{\eta}}\textbf{I}_r)$, where the hyperparameter $\tau^2_{\bm{\eta}}$ is chosen to be sufficiently large such that the prior distribution is nearly non-informative.

\subsection{Adaptive MCMC}\label{sec:MCMC}

Implementation of an MCMC sampling scheme for the AXING-need model seems challenging due to the dimensionality and distribution of the random coefficients $c_{jk}$'s.
One important observation that significantly reduces the computational burden is that the $t$-distribution (for $c_{jk}$'s) belongs to the class of scale mixtures of Gaussians (SMOG) \citep{Andrews-74, West-87}. Thus, the coefficient $c_{jk}$ can be expressed as $\sqrt{V_{jk}}G_{jk}$, where
$V_{jk}\sim \mathcal{IG}(\nu/2, \nu\sigma_j^2/2)$
%$$c_{jk}|V_{jk} \overset{indep.}{\sim} \mathcal{N}(0, V_{jk}),$$
%$$V_{jk}|\nu, \sigma_j \overset{indep.}{\sim} \mathcal{IG}\left( \frac{\nu}{2}, \frac{\nu\sigma_j^2 }{2}\right),$$
with $\mathcal{IG}(\alpha, \beta)$ denoting an inverse gamma distribution, and $G_{jk} \sim \mathcal{N}(0, 1)$ independent of $V_{jk}$. Since $c_{jk}$'s are independent, the same holds for $V_{jk}$'s and $G_{jk}$'s. This representation leads to an efficient MCMC algorithm in which a Gibbs sampler is used.
%$$p(x|\alpha, \beta)=\frac{\beta^{\alpha}}{\Gamma(\alpha)}x^{-\alpha-1}\exp\left( -\frac{\beta}{x} \right) \quad \mbox{for } x>0.$$

Let $\textbf{V}$ denote the vector stacked by $V_{jk}$'s, and $\bm{\sigma}^2$ denote the vector consisting of $\sigma_{J_0}^2,\cdots, \sigma_J^2$. We use a Gibbs sampler to sample from $[\textbf{c}, \textbf{V}, \bm{\theta}|\textbf{Z}]$ so that the full conditional distributions of $\textbf{c}, \textbf{V}, \bm{\sigma}^2$ and $\tau^2$ all have closed forms. In particular, the full conditional distribution of $\textbf{c}$ (i.e., $[\textbf{c}|\textbf{Z}, \textbf{V}, \bm{\theta}]$) is multivariate Gaussian. Sampling from it requires $\mathcal{O}(p^3)$ operations with $p=\sum_j p_j$, which is computationally intractable for large $p$. Nonetheless, our numerical experiments indicate that the subblocks $[\textbf{c}_j|\textbf{Z}, \textbf{V}, \bm{\theta}]$, $j=J_0,\cdots, J$ are weakly correlated, where $\textbf{c}_j=(c_{j1},\cdots, c_{jp_j})^{\rm T}$. This can be attributed to the fact that needlets are spatially localized and bandlimited. Thus, the sampling step for $\textbf{c}$ is achieved through successive draws from the conditional subblocks $[\textbf{c}_j|\textbf{Z}, \textbf{V}, \bm{\theta}, \textbf{c}_{-j}], j=J_0,\cdots, J$, where $\mathbf{c}_{-j}$ denotes the vector obtained by dropping $\textbf{c}_j$ from $\textbf{c}$. Partitioning the vector $\textbf{c}$ into subblocks avoids the Cholesky decomposition of  a large-scale matrix, while the weak correlation among pairs of the subblocks mitigates the potential problem of slow convergence for the Gibbs sampler. For sufficiently large $j$, we may further speed up the computation by sampling from $[c_{jk}|\textbf{Z}, \textbf{V}, \bm{\theta}, \textbf{c}_{-jk}], k=1,\cdots, p_j$. The full conditional distribution of $\bm{\eta}_{-0}$ is not available in closed form. Thus, we sample from $[\bm{\eta}_{-0}|\mathbf{Z}, \mathbf{c}, \mathbf{V}, \bm{\sigma}^2, \tau^2]$ using an adaptive Metropolis step \citep[Algorithm 4]{Andrieu-08}, and incorporate it into the Gibbs sampler.
 
Let $\textbf{A}_j$ denote the columns of $\textbf{A}$ corresponding to $j$, and $\textbf{V}_j=(V_{j1}, \cdots, V_{jp_j})^{\rm T}$. Then, the aforementioned adaptive Metropolis-within-Gibbs sampler is summarized as follows:
\begin{myalg}
\begin{enumerate}
\item Sample $\textbf{c}_j$ from $[\textbf{c}_j|\textbf{Z}, \textbf{V}, \bm{\theta}, \textbf{c}_{-j}]=\mathcal{N}(\hat{\bm{\mu}}_j, \widehat{\bm{\Sigma}}_j)$,
where 
$$\widehat{\bm{\Sigma}}_j=\tau^2\left(\textbf{A}_j^{\rm T}\textbf{G}^2\textbf{A}_j+\tau^2\mbox{diag} (\textbf{V}_j)^{-1} \right)^{-1},$$
and
$$\hat{\bm{\mu}}_j=\frac{1}{\tau^2} \widehat{\bm{\Sigma}}_j \textbf{A}_j^{\rm T} \textbf{G} (\textbf{Z}-\textbf{G}\textbf{A}_{-j}\textbf{c}_{-j}).$$
\item Sample $\textbf{V}$ from $[\textbf{V}|\textbf{Z}, \textbf{c}, \bm{\theta}]$, where $V_{jk}|\textbf{Z}, \textbf{c}, \bm{\theta}$ are independent and distributed as 
$$\mathcal{IG}\left(\frac{\nu+1}{2}, \frac{c_{jk}^2+\nu \sigma_j^2}{2} \right).$$
\item Sample $\bm{\sigma}^2$ from $[\bm{\sigma}^2|\textbf{Z}, \textbf{c}, \textbf{V}, \tau^2, \bm{\eta}]$, where $\sigma_j^2|\textbf{Z}, \textbf{c}, \textbf{V}, \tau^2, \bm{\eta}$
are independent and distributed as 
$$\mathcal{G}\left( \frac{\nu p_j}{2}, \frac{\nu}{2} \sum \limits_{k=1}^{p_j} \frac{1}{V_{jk}} \right),$$
where $\mathcal{G}(\alpha, \beta)$ denotes a gamma distribution.
\item Sample $\tau^2$ from 
$$[\tau^2|\textbf{Z}, \textbf{c}, \textbf{V}, \bm{\sigma}^2, \bm{\eta}]=\mathcal{IG}\left( \frac{n}{2}, \frac{(\textbf{Z}-\textbf{G}\textbf{A}\textbf{c})^{\rm T}(\textbf{Z}-\textbf{G}\textbf{A}\textbf{c})}{2} \right).$$
\item Sample $\bm{\eta}_{-0}$ using the adaptive Metropolis step from
$$[\bm{\eta}_{-0}|\textbf{Z}, \textbf{c}, \textbf{V}, \bm{\sigma}^2, \tau^2]\propto \\ \exp\left\{ -\frac{1}{2\tau^2}(\textbf{Z}-\textbf{G}\textbf{A}\textbf{c})^{\rm T}(\textbf{Z}-\textbf{G}\textbf{A}\textbf{c}) \right\}\exp \left\{ -\frac{1}{2\tau^2_{\bm{\eta}}}\bm{\eta}_{-0}^{\rm T}\bm{\eta}_{-0} \right\}.$$
The proposal distribution is chosen as
$$Q(\bm{\eta}_{-0}^*|\bm{\eta}_{-0}) \sim \mathcal{N}(\bm{\eta}_{-0}, \gamma \bm{\Sigma}),$$
where $\gamma$ is a parameter adaptively tuned with the goal of achieving the optimal acceptance rate \citep{Gelman-96}, and $\bm{\Sigma}$ is adaptively updated to approximate the covariance matrix of the full conditional distribution of $\bm{\eta}_{-0}$.
\end{enumerate}
\end{myalg}
Choosing a good initial value of the parameter vector $\bm{\theta}$ is crucial for successful convergence of the algorithm. We provide a simple but effective method to specify the initial value through fitting the model with $c_{jk} \sim \mathcal{N}(0, \nu\sigma_j^2/(\nu-2))$ for all $j,k$.  The MLE under this misspecified model, which assumes a Gaussian process for the observations, is easily computable, and close to the true value according to our numerical experiments. Thus, it is reasonable to specify the initial value of  $\bm{\theta}$ 
in terms of the MLE under the misspecified model. By default, the initial values of $\mathbf{c}$ and $\mathbf{V}$ are specified as a zero vector and an all-ones vector, respectively.

\subsection{Spatial prediction}\label{bayes_sp}

One main focus of spatial statistics is to predict values at unobserved locations based on observed values. In this subsection, we present a spatial prediction method for the proposed AXING-need model. Let $\mathbf{Z}^*=(Z(\mathbf{t}_1), \cdots, Z(\mathbf{t}_{n_{\rm P}}))^{\rm T}$, where $\mathbf{t}_1,\cdots, \mathbf{t}_{n_{\rm P}}$ are $n_{\rm P}$ unobserved locations. Then we have $\mathbf{Z}^*=\mathbf{G}^*\mathbf{A}^*\mathbf{c}+\mathbf{e}^*$, where $\mathbf{G}^*$, $\mathbf{A}^*$ and $\mathbf{e}^*$ are $\mathbf{G}, \mathbf{A}$ and $\mathbf{e}$ evaluated at the unobserved locations, respectively. From a fully Bayesian perspective, the posterior predictive distribution is
\begin{equation}\label{two_steps}
[\mathbf{Z}^*|\mathbf{Z}=\mathbf{z}]=\int_{\bm{\theta}, \mathbf{c}} [\mathbf{Z}^*|\bm{\theta}, \mathbf{c}, \mathbf{Z}=\mathbf{z}][\bm{\theta}, \mathbf{c}|\mathbf{Z}=\mathbf{z}]d\bm{\theta}d\mathbf{c},
\end{equation}
where $\mathbf{z}$ is the actual observed value of $\mathbf{Z}$, and
\begin{equation}\label{cond_Z}
\mathbf{Z}^*|\bm{\theta}, \mathbf{c}, \mathbf{Z}=\mathbf{z} \sim \mathcal{N}(\mathbf{G}^*\mathbf{A}^*\mathbf{c}, \tau^2 \mathbf{I}_{n_{\rm P}}).
\end{equation}
The posterior predictive distribution does not have a closed form, but it can be sampled using the MCMC samples from $[\bm{\theta}, \mathbf{c}|\mathbf{Z}=\mathbf{z}]$. Let  $\{ \bm{\theta}^{(l)}, \mathbf{c}^{(l)}, l=1,\cdots, L\}$ denote the MCMC samples. For each $l$, we draw a sample ${\mathbf{Z}^*}^{(l)}$ from $[\mathbf{Z}^*|\bm{\theta}^{(l)}, \mathbf{c}^{(l)}, \mathbf{Z}=\mathbf{z}]$ in (\ref{cond_Z}).
The empirical distribution of $\{ {\mathbf{Z}^*}^{(l)}, l=1,\cdots, L\}$ is approximately the posterior predictive distribution.
Thus, the posterior predictive mean, standard deviation, quantiles and posterior predictive intervals can be computed based on these samples, respectively.
For example, the posterior predictive mean is estimated by $L^{-1}\sum_{l=1}^L {\mathbf{Z}^*}^{(l)}$, and the $(1-\alpha)100\%$ posterior predictive interval is estimated by finding the $(100\alpha/2)$-th and $(100(1-\alpha/2))$-th percentiles of $\{ {\mathbf{Z}^*}^{(l)}, l=1,\cdots, L\}$.

\subsection{Unconditional simulation}

Utility of spatial models is not restricted to spatial prediction. 
For example, \citet[Section 7.4]{Genton-15} emphasized the importance of using spatial models for simulation of random fields.  It is also well-known that a misspecified spatial model does not necessarily mean significantly inferior predictive performance \citep{Stein-88}. Thus, simulation of random fields from a fitted spatial model provides an alternative way to test the performance of the model. Specifically, it is used to check whether the proposed AXING-need model can successfully capture the characteristics of data. The following algorithm describes the details of simulating a random field from a fitted AXING-need model (\ref{AXING_matrix}) at locations $\mathbf{s}_1, \cdots, \mathbf{s}_{n_{\rm S}}$.
\begin{myalg}\label{alg_sim}
\begin{enumerate}
\item To incorporate the uncertainty of parameter estimates into the simulation, we randomly select one of the MCMC samples $\{ \bm{\theta}^{(l)}, l=1,\cdots, L\}$, denoted by 
$\hat{\bm{\theta}}=(\hat{\sigma}_{J_0}^2, \cdots, \hat{\sigma}_{J}^2, \hat{\tau}^2, \hat{\bm{\eta}}_{-0})$,
as an estimate of the parameters. The matrix $\mathbf{G}$ is evaluated with $\bm{\eta}_{-0}=\hat{\bm{\eta}}_{-0}$ at the locations $\mathbf{s}_1, \cdots, \mathbf{s}_{n_{\rm S}}$, and the design matrix $\mathbf{A}$ is also evaluated at these locations.
\item Generate a coefficient vector $\mathbf{c}$, where $c_{jk} \sim \hat{\sigma}_j t(\nu)$ for all $j, k$, and they are independent.
\item Then, $\mathbf{X}=\mathbf{G}\mathbf{A}\mathbf{c}$ is an unconditional simulation of random fields from the fitted AXING-need model, where the term of observational errors is omitted.
\end{enumerate}
\end{myalg}

\section{Numerical experiments}\label{sec:num_exp}
\subsection{Accuracy of parameter estimation}\label{sec:acc_param}

In this subsection, we investigate the accuracy of parameter estimation by a Monte Carlo simulation study. The data are generated from the proposed AXING-need model  (\ref{AXING_matrix}), and the sampling locations are on a mildly perturbed HEALPix grid \citep{Gorski-05} with 768 grid points. To reduce the computational burden of the simulation study, we assume that there are only two levels of needlets, i.e., $J_0=2, J=3$. As a function of co-latitude, $\log{g}$ is represented as a linear combination of cubic B-splines with one interior knot $\pi/2$, where the first B-spline is replaced by the intercept. We consider two different settings of the coefficient vector $\bm{\eta}_{-0}$, and specify $\eta_0=0, \nu=4, \sigma_2=1.25, \sigma_3=0.4419, \tau=0.1$.

\begin{figure}[htbp] %  figure placement: here, top, bottom, or page
   \centering
   \begin{tabular}{cc}
		\hspace{-0.1in} \includegraphics[width = 0.5\linewidth]{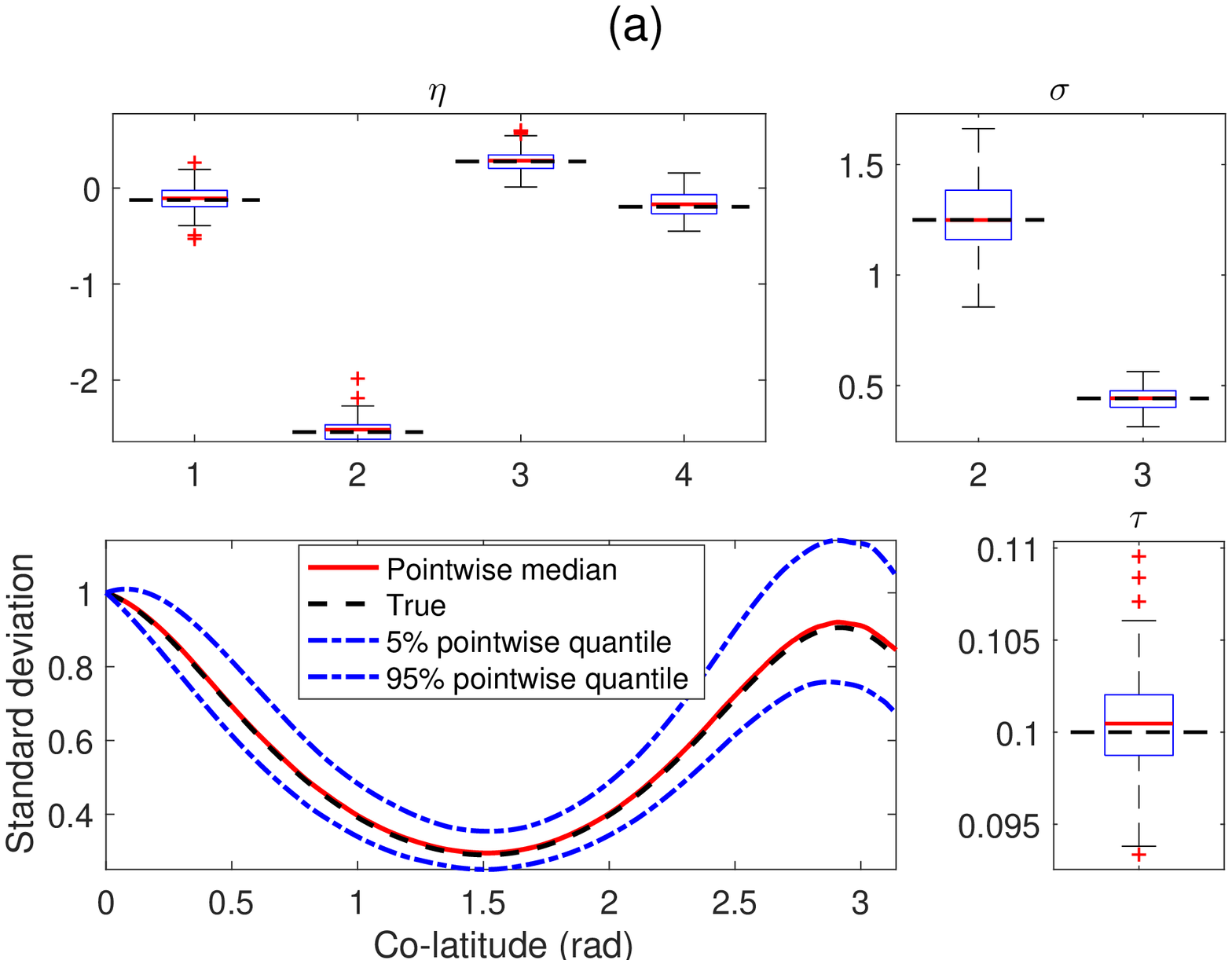} \hspace{-0.2in} & \includegraphics[width = 0.5\linewidth]{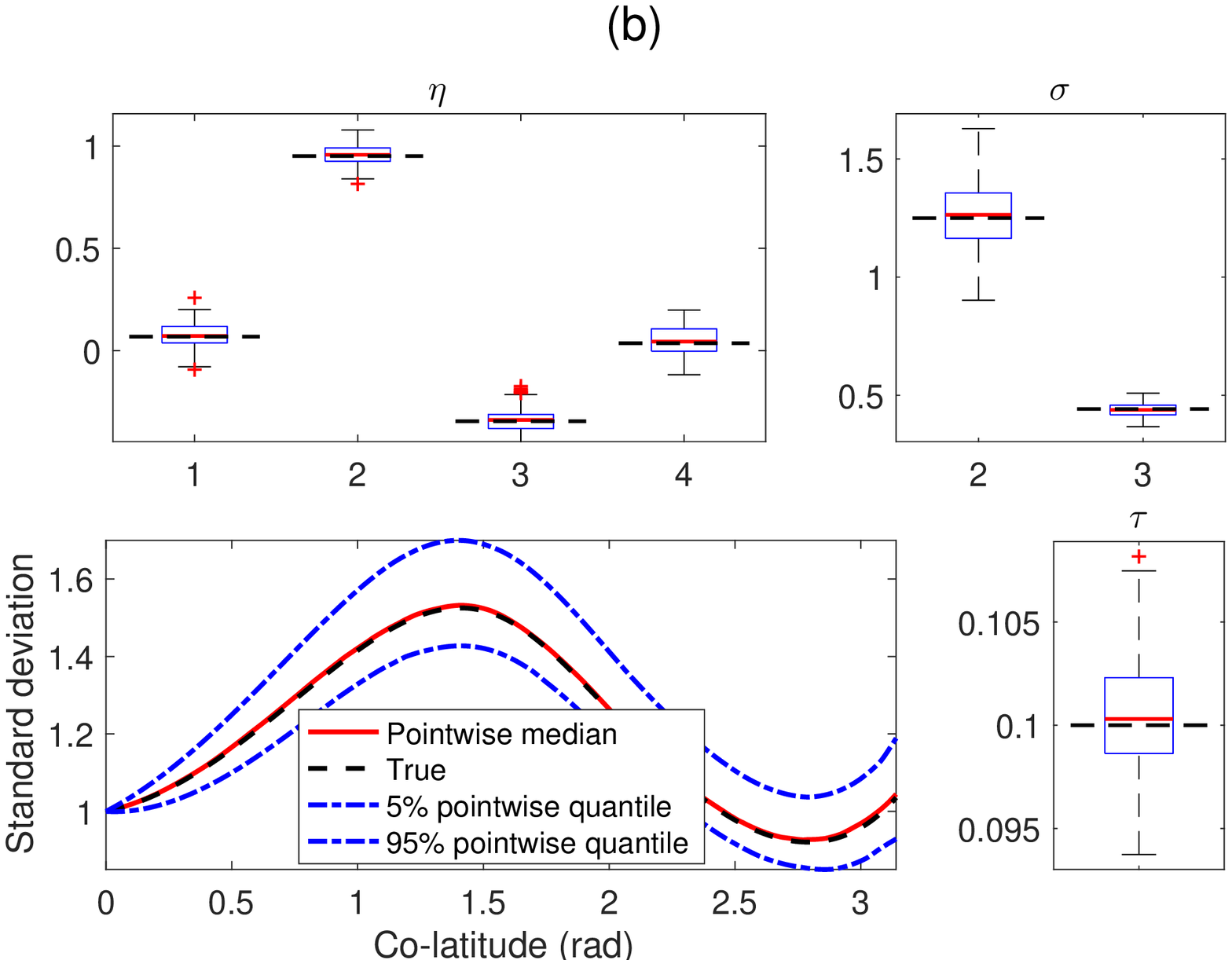}\\
   \end{tabular}
   \caption{Results of the Monte Carlo simulation study under two different settings of $\bm{\eta}_{-0}$ in \textbf{panels (a)} and \textbf{(b)}. The dashed horizontal line accompanying the boxplots shows the true value of the parameters. \textbf{Top left block:} boxplots of estimated $\bm{\eta}_{-0}$; \textbf{Bottom left block:} estimated and true curves of $g$ (i.e., the standard deviation profile up to a constant); \textbf{Top right block:} boxplots of estimated $\sigma_2$ and $\sigma_3$; \textbf{Bottom right block:} boxplot of estimated $\tau$.}
   \label{fig:sim_para_acc_1}
\end{figure}

For each simulation run, the parameters $(\bm{\eta}_{-0}, \sigma_2, \sigma_3, \tau)$ are estimated by the posterior sample means using the adaptive MCMC algorithm in Section \ref{sec:MCMC}, in which 
$\tau_{\bm{\eta}}=10$. The chain was run for 400,000 iterations, with the first 200,000 samples discarded as a burn-in period, and every $200$th sample is collected to compute the posterior sample means.
Figure \ref{fig:sim_para_acc_1} displays the boxplots of the parameter estimates based on $100$ simulation runs. We can see that all the parameter estimates have small biases and low variability. Moreover, the pointwise median curve of the function $g$ matches well with  the true one. These demonstrate the effectiveness of the estimation procedure using the adaptive MCMC algorithm.

\subsection{Predictive performance comparison}\label{sec:pred_perf_comp}

\begin{figure}[htbp] %  figure placement: here, top, bottom, or page
   \centering
   \begin{tabular}{ccc}
		\hspace{-0.15in} \includegraphics[width= 0.35 \linewidth]{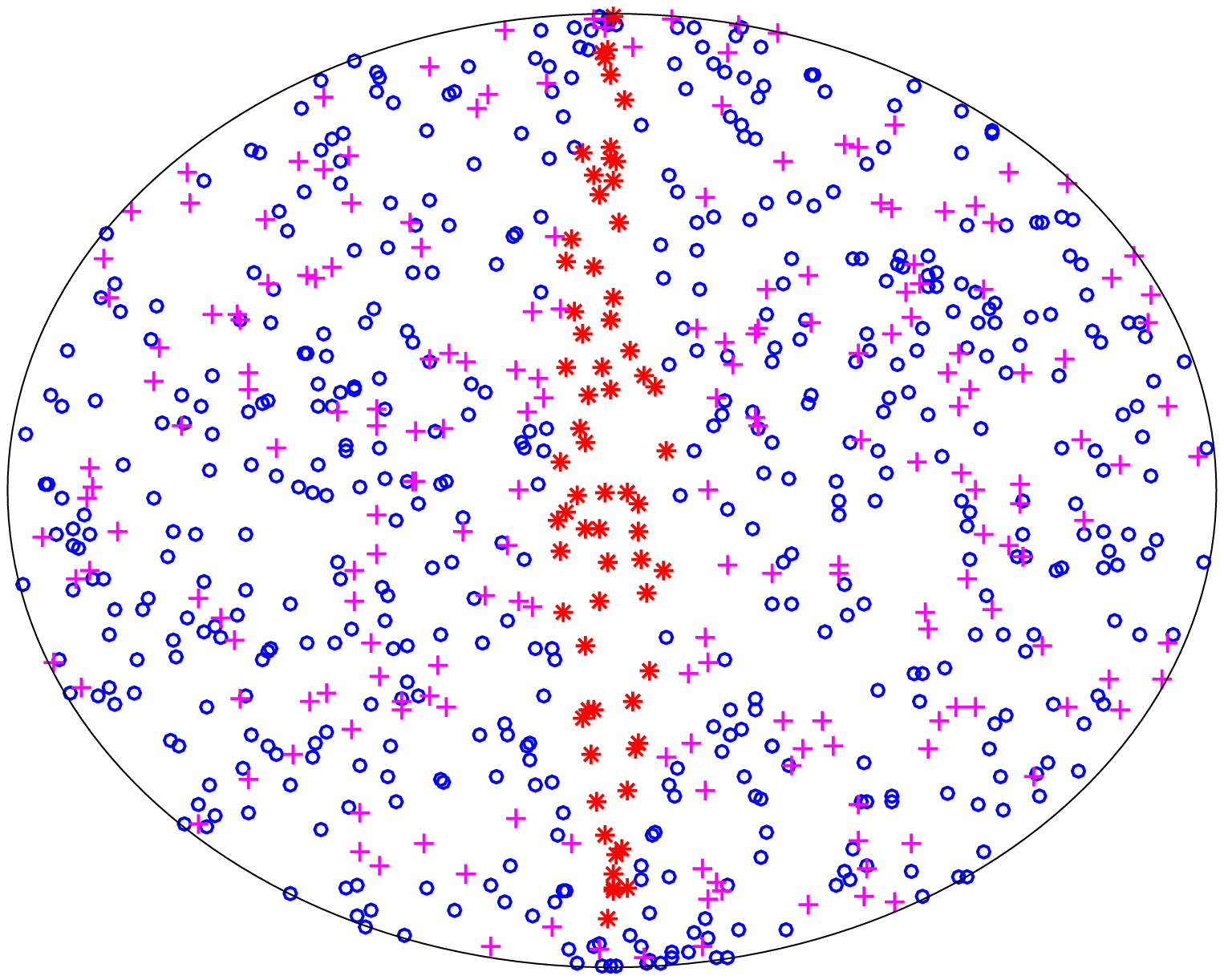}  \hspace{-0.35in} & \includegraphics[width = 0.35\linewidth]{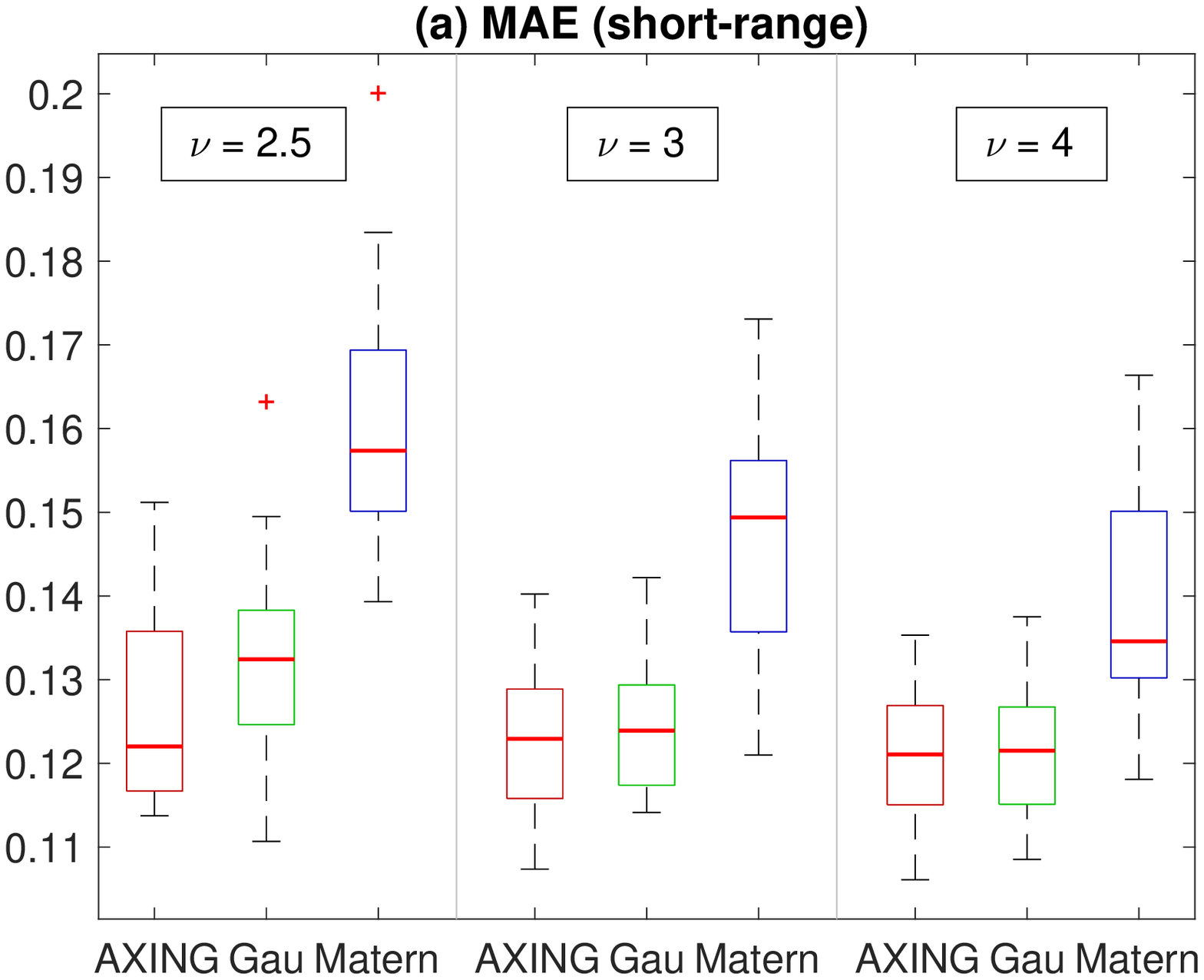} \hspace{-0.35in} & \includegraphics[width = 0.35\linewidth]{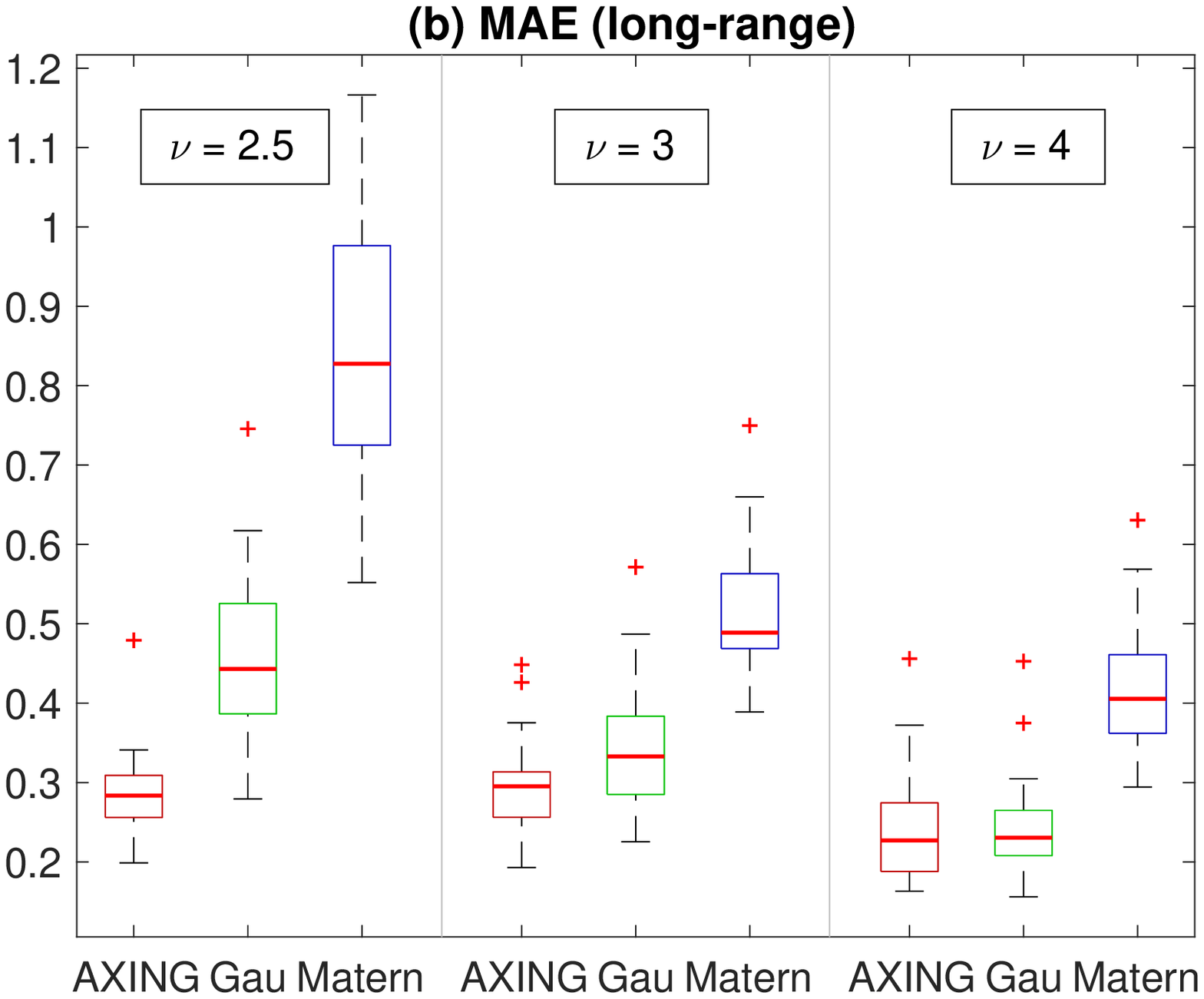} \\
		\hspace{-0.15in} \includegraphics[width = 0.35\linewidth]{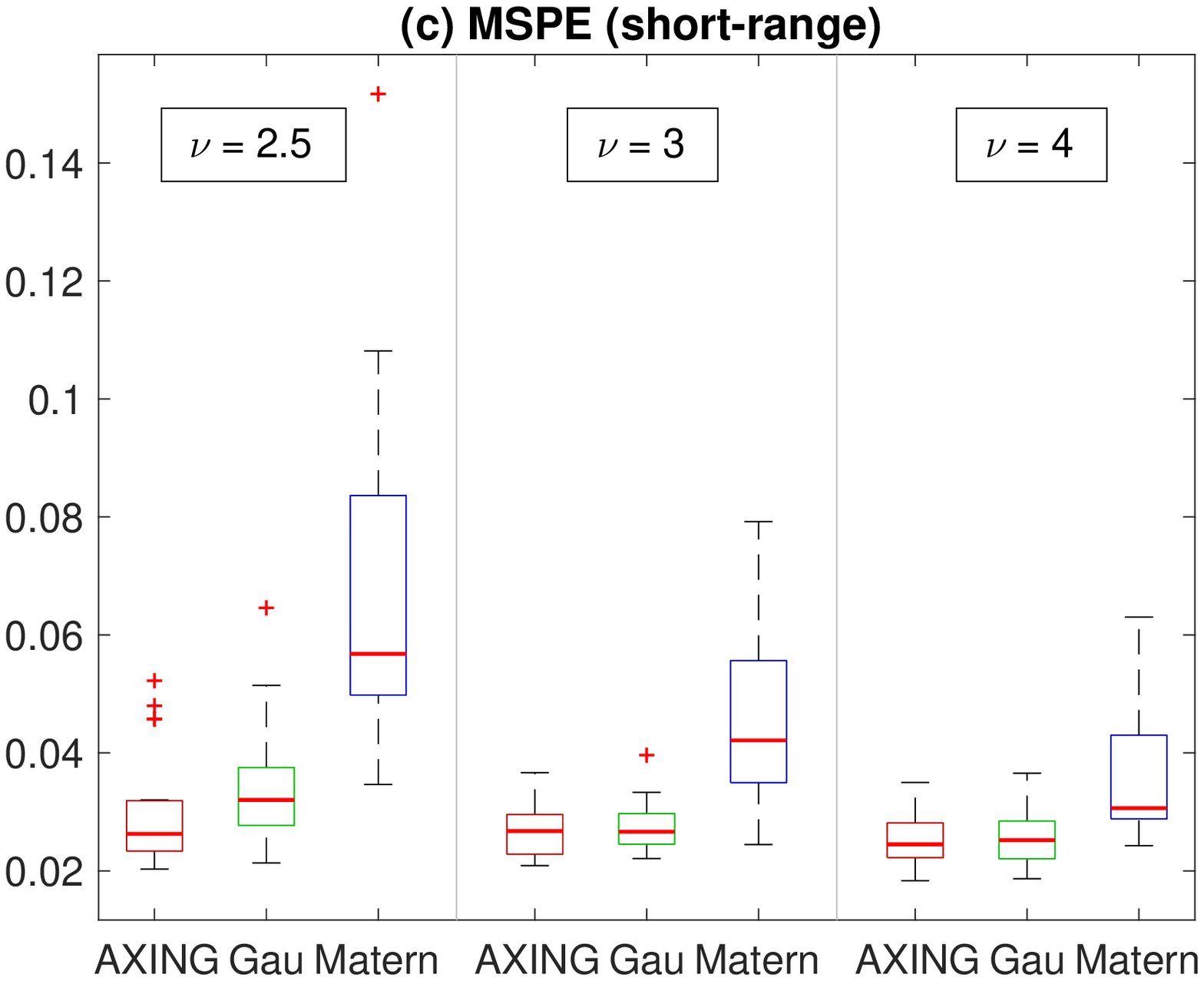} \hspace{-0.35in} & \includegraphics[width = 0.35\linewidth]{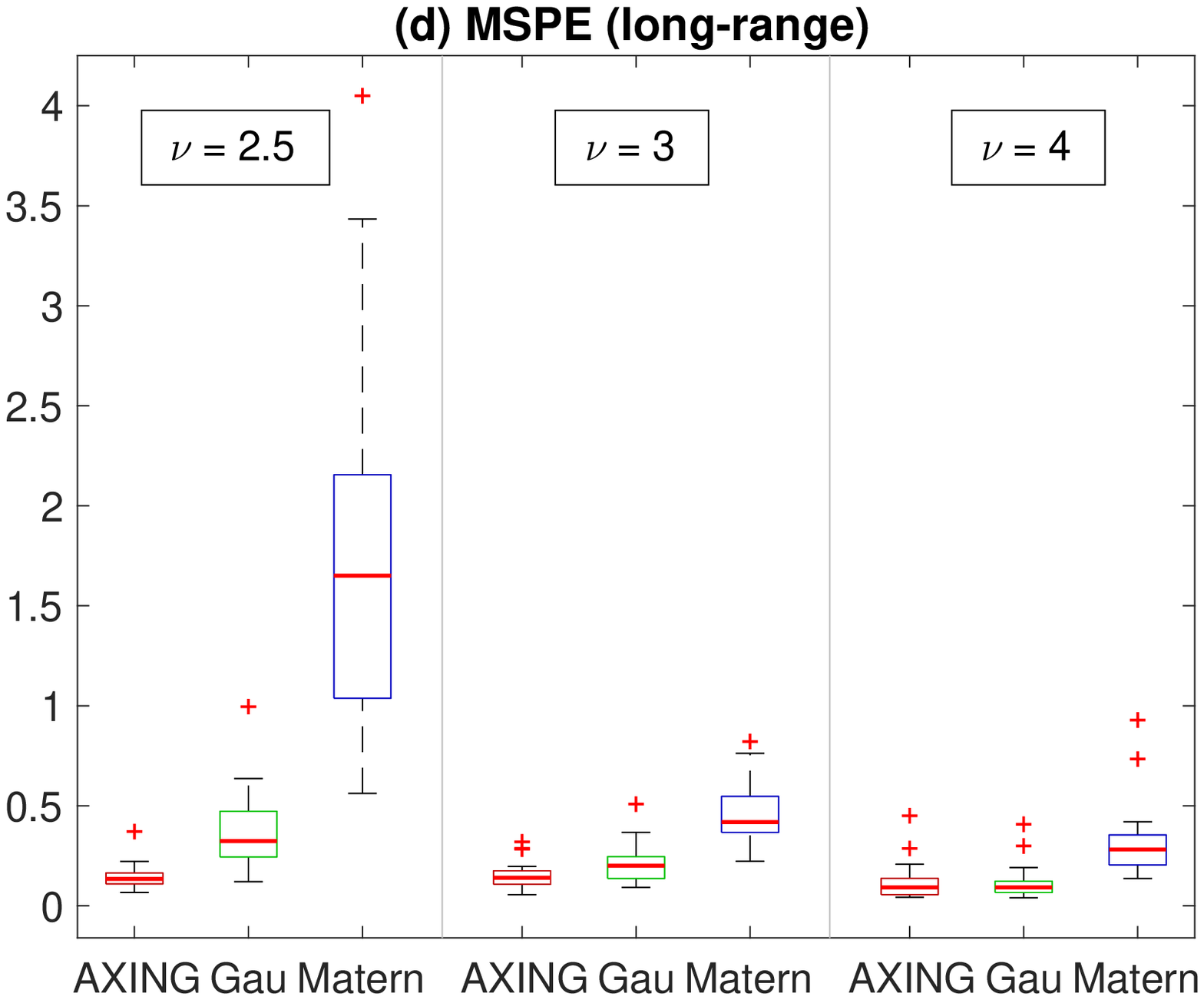} \hspace{-0.35in} & \includegraphics[width = 0.35\linewidth]{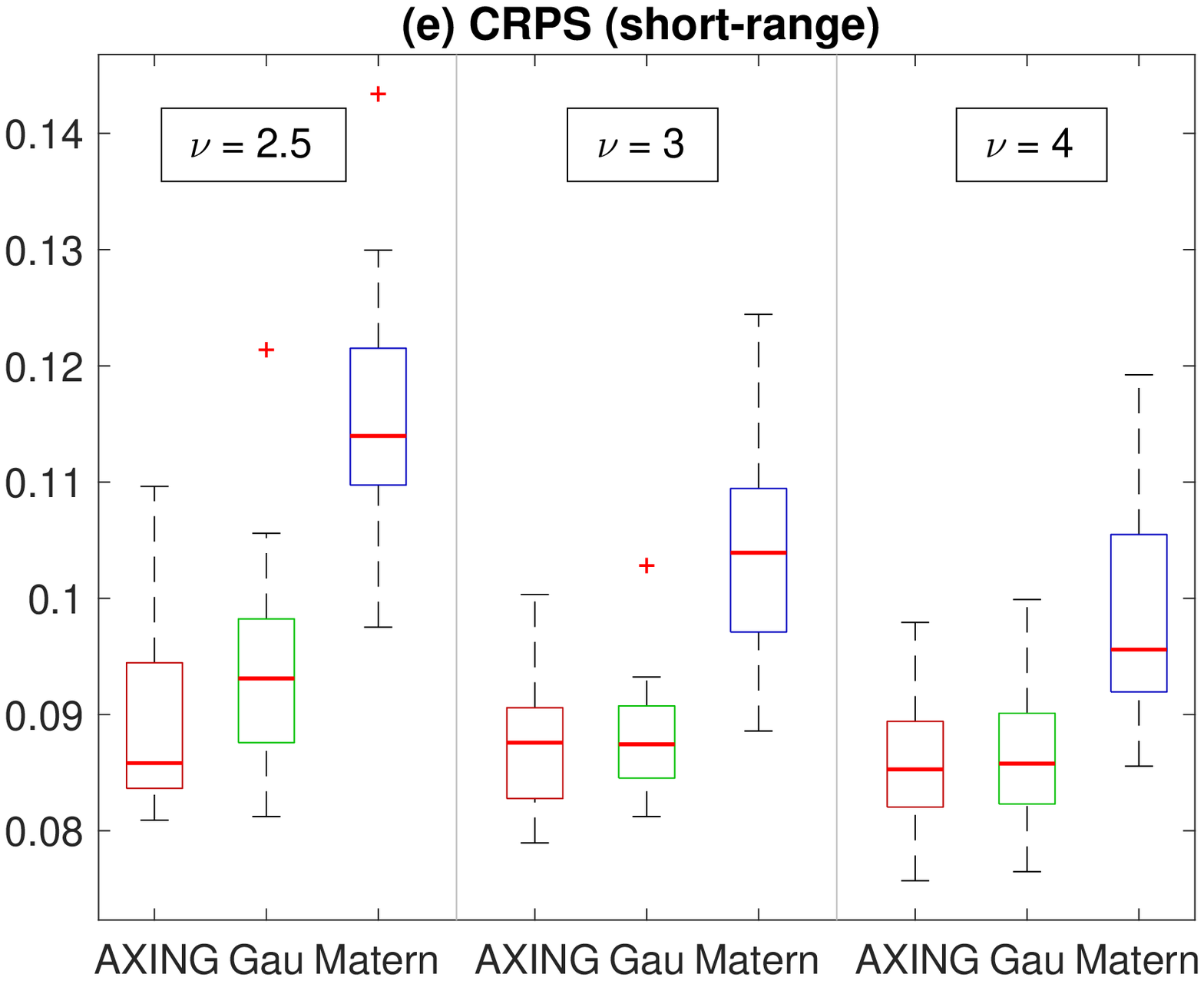} \\
		\hspace{-0.15in} \includegraphics[width = 0.35\linewidth]{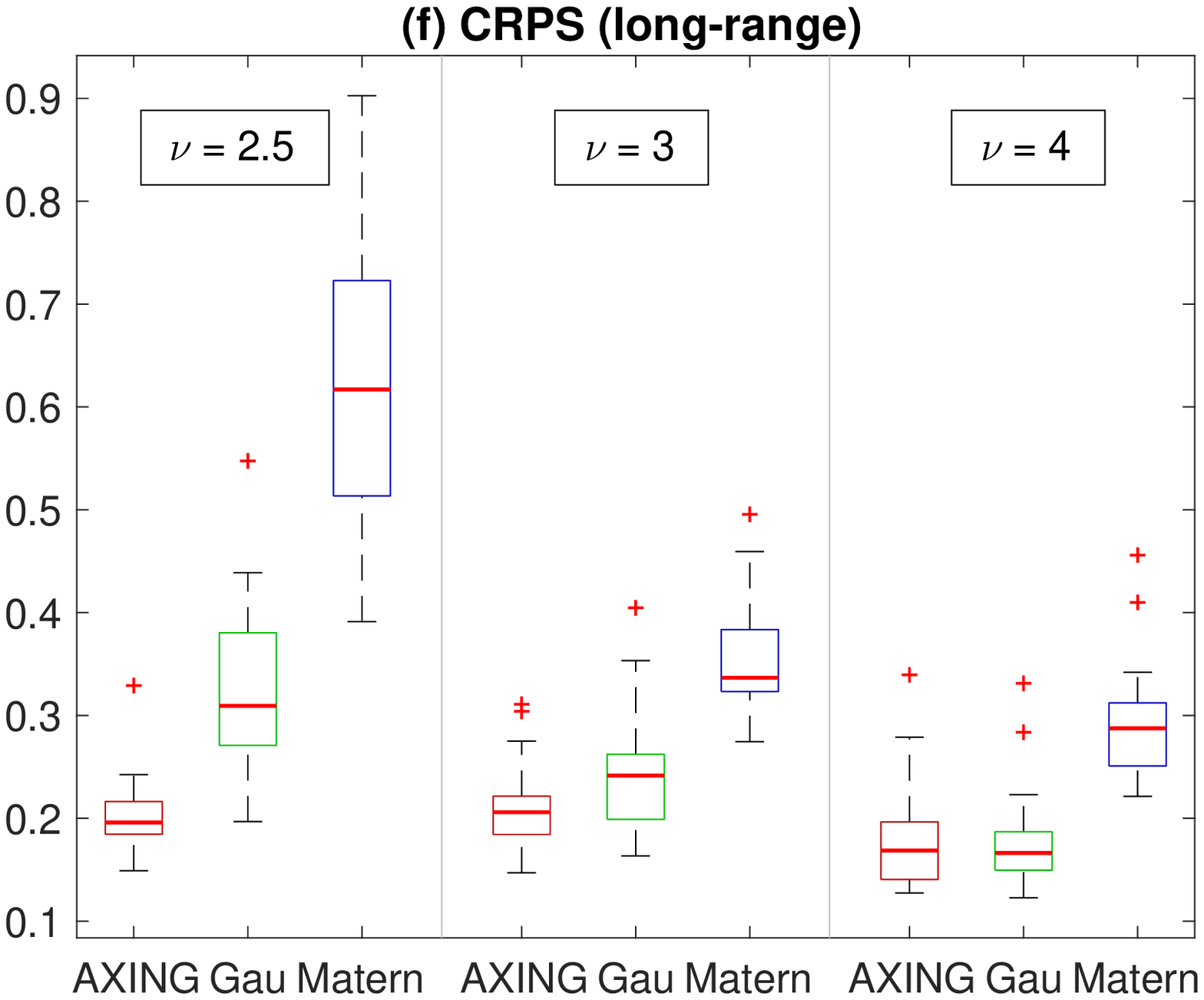} \hspace{-0.35in} & \includegraphics[width = 0.35\linewidth]{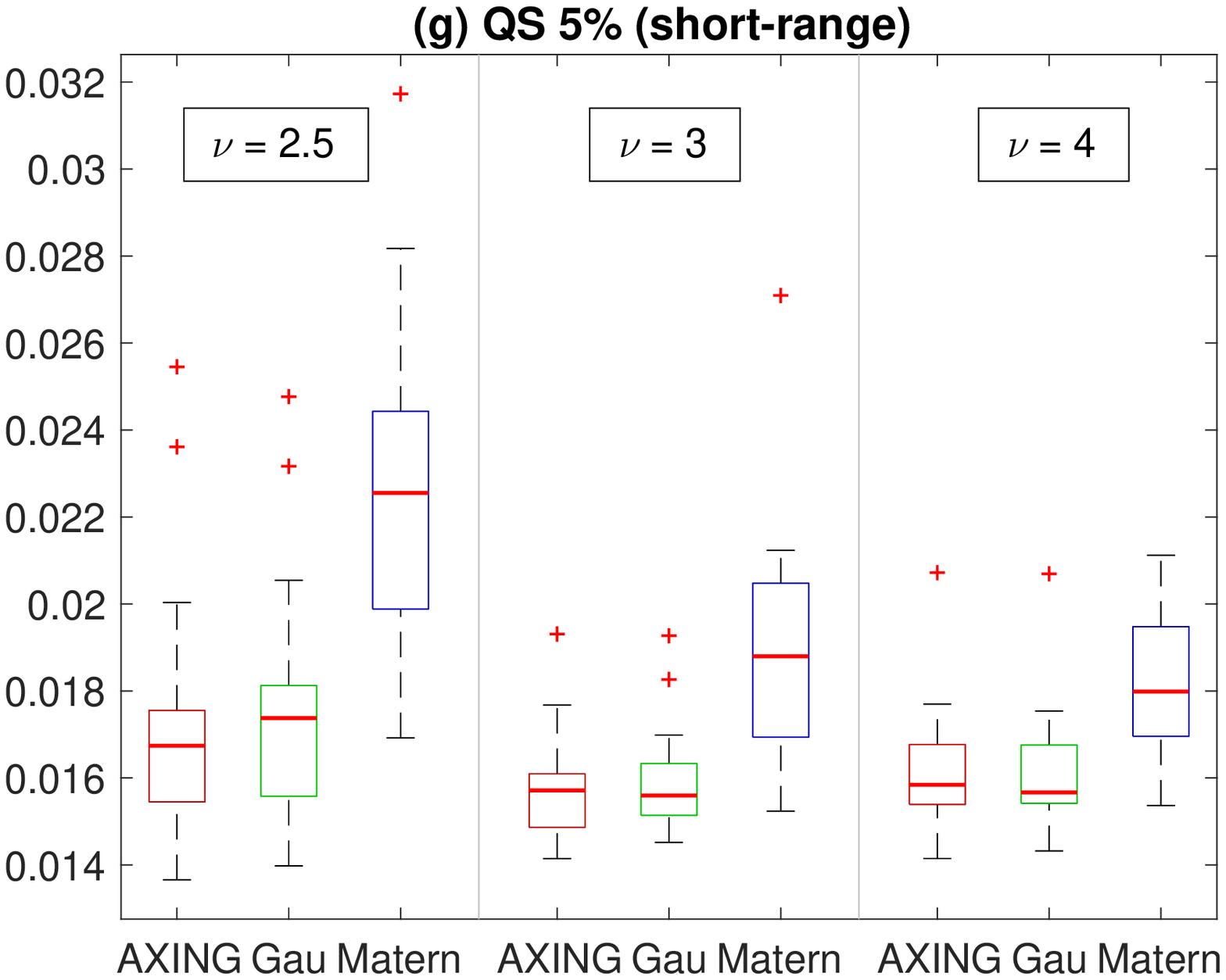} \hspace{-0.35in} & \includegraphics[width = 0.35\linewidth]{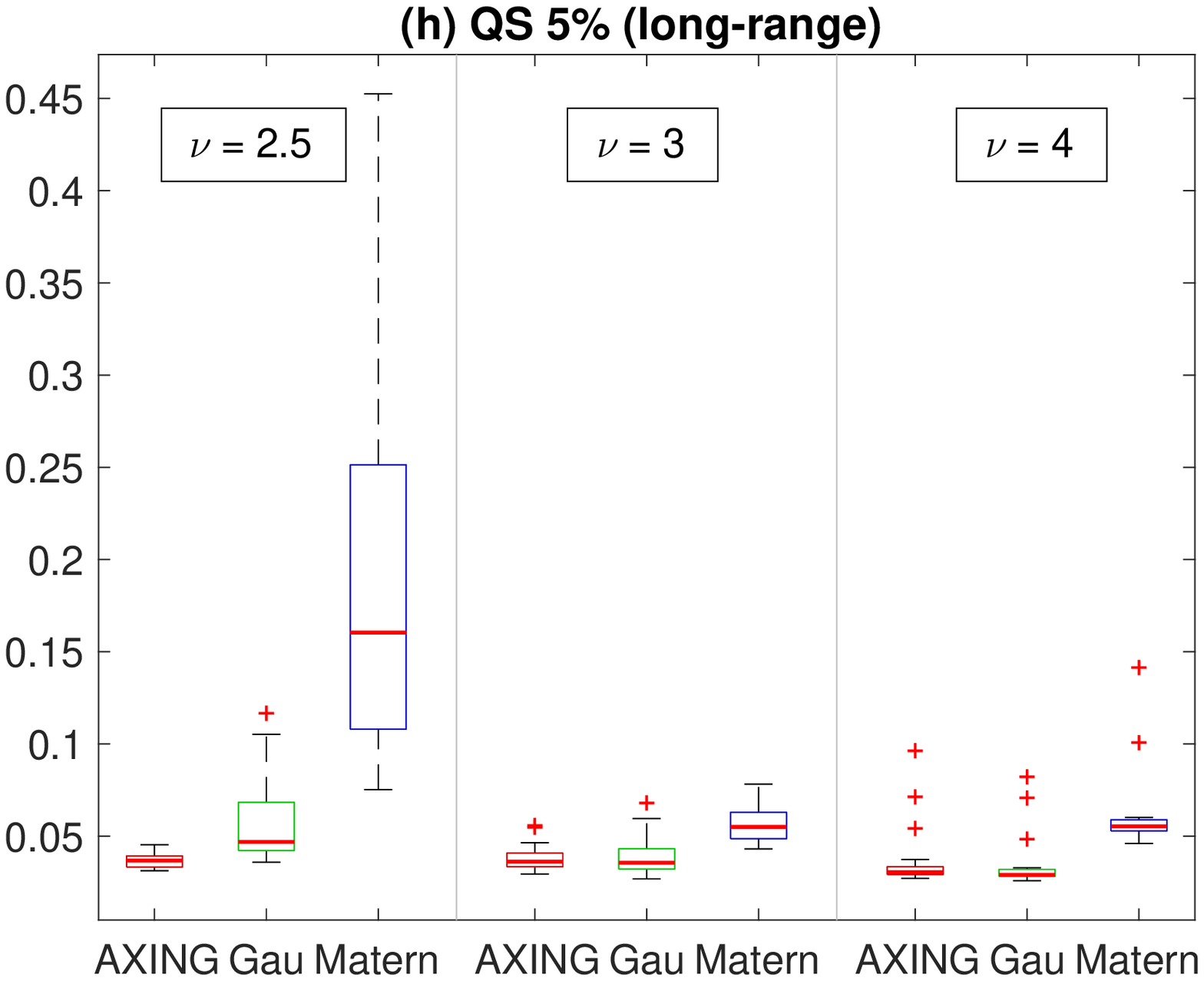} \\
		\hspace{-0.15in} \includegraphics[width = 0.35\linewidth]{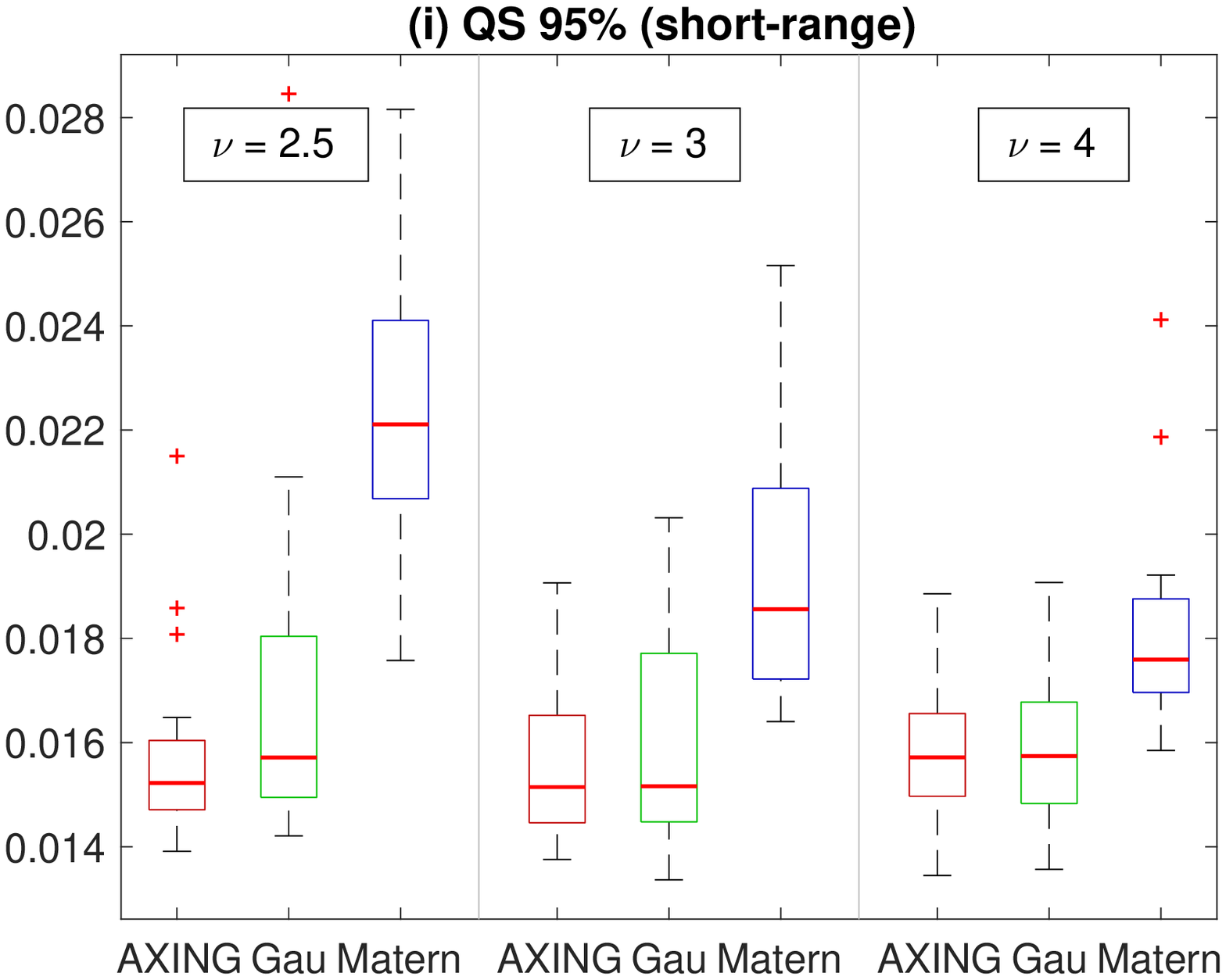} \hspace{-0.35in} & \includegraphics[width = 0.35\linewidth]{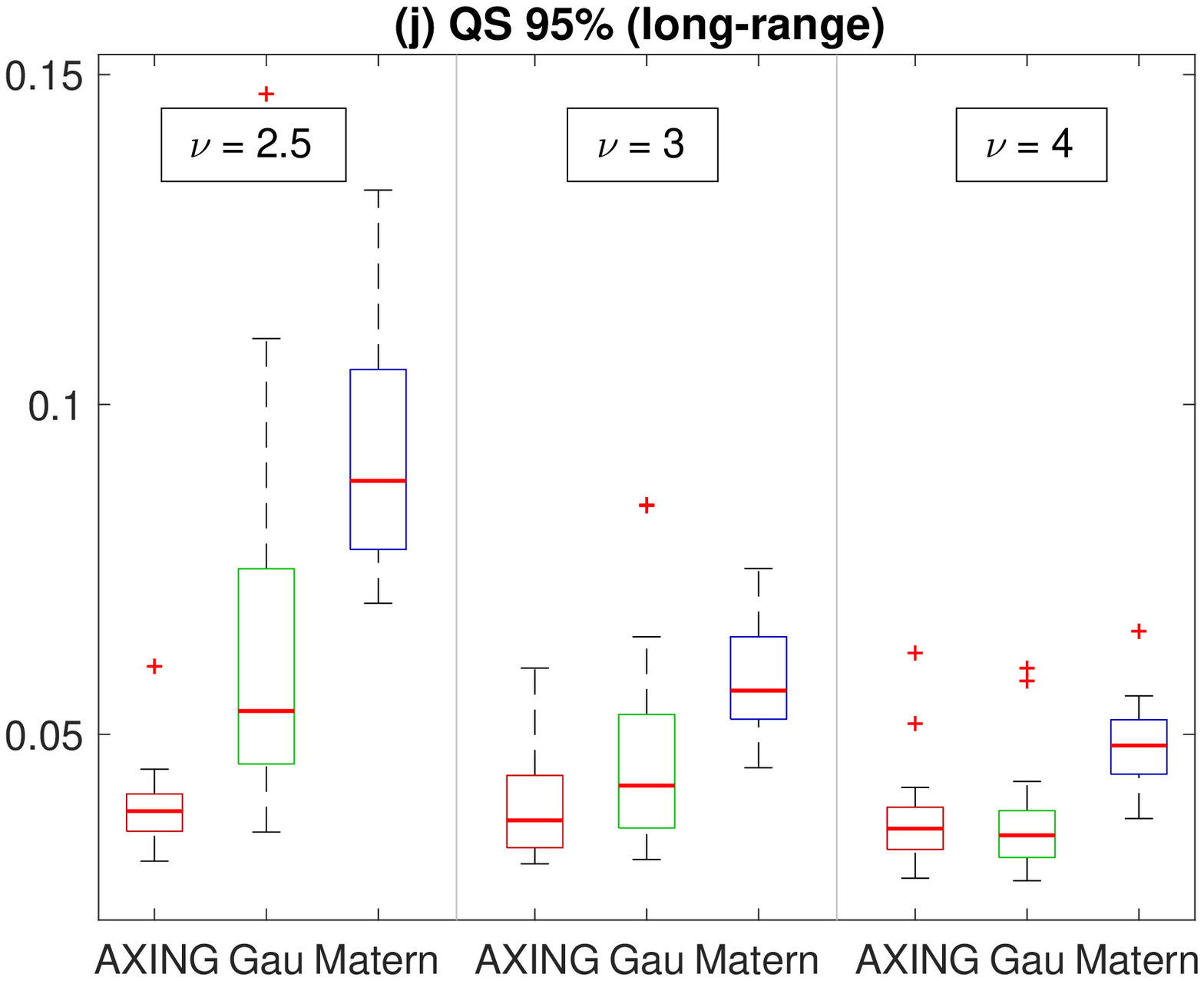} \hspace{-0.35in} & 
   \end{tabular}
   \caption{\textbf{First panel:} Training and test set separation for one of the cross-validation replications. The sphere has been projected to an ellipse by the Hammer projection. ``$\circ$" observed locations; ``$+$" short-range predicted locations; ``$\ast$" long-range predicted locations; \textbf{Other panels:} Boxplots of five scoring rules for the AXING-need, Gau-need and Gau-Mat\'ern models with $20$ cross-validation replications. \textbf{(a)-(b)} MAE; \textbf{(c)-(d)} MSPE; 
   \textbf{(e)-(f)} CRPS; \textbf{(g)-(h)} QS 5\%; \textbf{(i)-(j)} QS 95\%.}
   \label{fig:pred_perf}
\end{figure}

In this subsection, we compare the predictive performance of the proposed AXING-need model with that of two Gaussian models, which are \textit{Gaussian needlet (Gau-need)} and \textit{Gaussian Mat\'ern (Gau-Mat\'ern)} models, the latter being widely used in spatial statistics. The purpose of this comparison is to understand the effects of two factors on spatial predictive performance: Gaussian versus non-Gaussian, and 
a Mat\'ern covariance structure versus an oscillating one constructed through a needlet representation of the process. In the Gau-need model, the coefficient vector $\textbf{c}$ in (\ref{AXING_matrix}) is assumed to be multivariate Gaussian, i.e., 
$$\textbf{c} \sim \mathcal{N}(\textbf{0}, \bm{\Lambda}),$$
where $\bm{\Lambda}$ is a diagonal matrix, and $\mbox{Var}(c_{jk})=\sigma_j^2$ for all $j, k$.
The Gau-Mat\'ern model assumes a nonstationary Mat\'ern covariance function with latitude-dependent variance
$$C(\textbf{s}, \textbf{t})=g(\textbf{s})g(\textbf{t})M(\lVert \textbf{s}-\textbf{t} \rVert; \kappa, a),$$
where $\textbf{s}, \textbf{t}\in \mathbb{S}^2$, $\lVert \textbf{s}-\textbf{t} \rVert$ denotes the chordal distance between the two locations $\textbf{s}$ and $\textbf{t}$, and $g$ is given by (\ref{gs}). Besides, $M(r; \kappa, a)$ is the Mat\'ern correlation function
$$M(r; \kappa, a)=\frac{2^{1-\kappa}}{\Gamma(\kappa)}(ar)^{\kappa}K_{\kappa}(ar),$$
where $K_\kappa$ is the modified Bessel function of the second kind, $\Gamma$ is the gamma function, $\kappa > 0$ is the smoothness parameter, and $a > 0$ is the spatial scale parameter, whereby $1/a$ controls the range of correlation.

The data are generated from the AXING-need model, where the parameter $\nu$ is specified as $2.5, 3, 4$ to illustrate the impact of the degree of non-Gaussianity on prediction accuracy. The specification of the remaining model parameters and sampling locations are the same as the first setting in Section \ref{sec:acc_param}. For each cross-validation replication, we randomly select $500$ locations outside a fixed longitudinal region with width $30^\circ$ as the training set to estimate the parameters. The remaining locations are held out as the test set to evaluate the predictive performance. For the AXING-need model, the parameters are estimated by the adaptive MCMC algorithm described in Section \ref{sec:MCMC}, in which $\tau_{\bm{\eta}}=10$. The chain was run for 1,000,000 iterations, with the first 500,000 samples discarded as a burn-in period, and every $500$th sample is collected as the posterior samples.
The posterior predictive mean and quantiles (see Section \ref{bayes_sp}) are used for prediction.
For the Gaussian models, the parameters are estimated by the maximum likelihood method, and the prediction is performed by the usual kriging. The first panel of Figure \ref{fig:pred_perf} shows the training and test set separation for one of the cross-validation replications. Held-out locations in and out of the longitudinal region are used to test the predictive performance in terms of long-range and short-range predictions, respectively. We assess the prediction accuracy using five scoring rules: the mean absolute error (MAE), the mean squared prediction error (MSPE), the continuous ranked probability score (CRPS) \citep{Gneiting-07}, and the quantile scores at the levels of 5\% and 95\% \citep{gneiting-11}, averaged over all the predicted locations. The CRPS for the AXING-need model is computed based on the MCMC samples (see Section S.1.3 of the supplementary material \citet{Fan-17-supp} for details). The quantile score for a quantile forecast $q$ at the level $\alpha \in (0, 1)$ is defined as $\mbox{QS}_{\alpha}(q, y)=(\mathbbm{1}\{ y<q \}-\alpha)(q-y)$, where $y$ is the observed value. We repeat the cross-validation procedure 20 times, and summarize the prediction accuracy of the models in terms of the scoring rules in the remaining panels of Figure \ref{fig:pred_perf}. We can see that when $\nu=3, 4$, the AXING-need and Gau-need models give similar results, and both are better than the Gau-Mat\'ern model. When $\nu=2.5$, the Gau-need model significantly deviates from the non-Gaussian characteristics of the data, and thus performs worse than the AXING-need model. Again, the Gau-Mat\'ern model has the worst predictive performance among the three. The difference in predictive performance between the Gau-need and Gau-Mat\'ern models becomes more pronounced for the long-range prediction than the short-range one. This can be explained by the following facts: the Gau-Mat\'ern model can reasonably approximate the true covariance structure at short distances, but it fails to capture the oscillating covariance structure at long distances due to its constraint of nonnegative covariance. 

\begin{table}[htp]
\caption{Coverage probabilities (CP) and mean lengths (mLen) of $50\%$ and $90\%$ predictive intervals, averaged over $20$ cross-validation replications}
\begin{center}
\resizebox{\textwidth}{!}{%
\begin{tabular}{c|c|cc|cc|cc|cc}
\hline
\hline
\multirow{2}{*}{$\nu$} & \multirow{2}{*}{Model} & \multicolumn{4}{|c|}{Short-range prediction} & \multicolumn{4}{|c}{Long-range prediction}\\
\cline{3-10}
& & CP($50\%$) & mLen & CP($90\%$) & mLen & CP($50\%$) & mLen & CP($90\%$) & mLen \\
\hline
\multirow{3}{*}{$2.5$} &  AXING-need & 49.4\% & 0.21 & 88.8\% & 0.50 & 53.2\% & 0.55 & 91.8\% & 1.34\\
& Gau-need & 48.3\% & 0.21 & 88.2\% & 0.51 & 34.9\% & 0.57 & 79.7\% & 1.39 \\
& Gau-Mat\'ern & 52.6\% & 0.28 & 90.6\% & 0.69 & 43.4\% & 1.06 & 81.7\% & 2.59 \\
\hline
\multirow{3}{*}{$3$} &  AXING-need & 49.0\% & 0.20 & 90.0\% & 0.50 & 50.5\% & 0.52 & 88.6\% & 1.26\\
& Gau-need & 48.2\% & 0.20 & 89.2\% & 0.49 & 44.0\% & 0.50 & 84.7\% & 1.22  \\
& Gau-Mat\'ern & 51.3\% & 0.25 & 90.6\% & 0.61 & 44.4\% & 0.87 & 93.2\% & 2.12 \\
\hline
\multirow{3}{*}{$4$} &  AXING-need & 50.7\% & 0.20 & 89.6\% & 0.50 & 63.5\% & 0.50 & 93.2\% & 1.23\\
& Gau-need & 50.4\% & 0.20 & 89.4\% & 0.49 & 59.5\% & 0.48 & 92.1\% & 1.16\\
& Gau-Mat\'ern & 52.1\% & 0.24 & 90.6\% & 0.60 & 54.3\% & 0.78 & 93.5\% & 1.90 \\
\hline
\end{tabular}}
\end{center}
\label{table:pred_sharpness}
\end{table}%

\citet{Gneiting-07-Sharpness} proposed the concept of \emph{sharpness} to evaluate the predictive performance in probabilistic prediction, which provides a predictive distribution instead of a point prediction for the value of a random field at an unobserved location. Sharpness is measured by the concentration of a predictive distribution around the true value, and the more concentrated the predictive distribution is, the sharper the prediction. We adopt the mean length of predictive intervals (``mean" refers to averaging over all the predicted locations) and the corresponding coverage probability  to assess the sharpness of the predictions produced by the models. The posterior predictive interval (see Section \ref{bayes_sp}) and the usual symmetric predictive interval are used for the AXING-need model and the Gaussian models, respectively. Table \ref{table:pred_sharpness} displays the mean lengths of $50\%$ and $90\%$ predictive intervals and the corresponding coverage probabilities, averaged over 20 cross-validation replications. In general, the AXING-need model yields the shortest predictive intervals with satisfactory coverage probabilities.

\section{Application to high-latitude ionospheric electrostatic potentials}\label{sec:app}

In this section, we demonstrate the effectiveness of the proposed AXING-need model and the associated statistical methodology through applying them to the LFM-MIX model output: the high-latitude ionospheric electrostatic potentials, which are introduced in Section \ref{sec:intro}. A detailed description of the LFM-MIX model can be found in \citet{Wiltberger-16}.
The LFM-MIX model is capable of running at multiple resolutions including Single, Double and Quad. Herein we use Quad resolution simulations since they contain more small-scale details than the other two.  At the Quad resolution, the sampling locations are on a regular $1\times 1$ degree geomagnetic latitude-longitude grid with latitudes ranging from $45^\circ$ to $90^\circ$. The electrostatic potentials are generated every two minutes during the time period from March 20, 2008 to April 16, 2008 (inclusive), and thus there are $20160$ time points in total. We denote the observations by $Z(\mathbf{s}_i, t_r), i=1, \cdots, N, r=1, \cdots, T$, where $\mathbf{s}_i$ and $t_r$ represent location and time, respectively.
%As shown in the Q-Q plots of Figure \ref{fig:real_nonGau}, the small-scale component of the electrostatic potentials exhibits a significant degree of non-Gaussianity. 

\subsection{Data preprocessing}\label{sec:data_preproc}
To extract the small-scale component of the electrostatic potentials, we subtract a crude estimate of the large-scale component from the observations. The large-scale component is estimated from the data by the empirical orthogonal function (EOF) method, which was applied to the LFM-MIX model output in \citet{kleiber-13}. Let $\mathbf{Z}=(Z(\mathbf{s}_i, t_r))_{1\leq r \leq T,1\leq i\leq N}$ denote the $T \times N$ matrix of observations with each row corresponding to a time point. We center $\mathbf{Z}$ by subtracting the vector of column averages from each row. The singular value decomposition (SVD) is then applied to $\mathbf{Z}$, i.e., $\mathbf{Z}=\mathbf{U}\mathbf{D}\mathbf{V}^{\rm T}$, where $\mathbf{U}=(u_{rk})_{1\leq r \leq T,1\leq k \leq T}$ is a $T\times T$ orthogonal matrix, $\mathbf{D}=\mbox{diag}(d_1, \cdots, d_T)$ with decreasing singular values $d_k$'s, and $\mathbf{V}=(v_{ik})_{1\leq i \leq N,1\leq k \leq T}$ is an $N\times T$ matrix with orthonormal columns, i.e., EOFs. Element-wise, we can express the observations as $Z(\mathbf{s}_i, t_r)=\sum_{k=1}^Td_ku_{rk}v_{ik}$. The cumulative sum of the squared singular values indicates that the first four EOFs can explain approximately 95\% of the total variability in the data. Section S.2.1 of the supplementary material \citep{Fan-17-supp} gives a scientific explanation of the EOFs through comparing them with those derived from real observed data. Thus, the large-scale component can be estimated by $\sum_{k=1}^K d_ku_{rk}v_{ik}$ with $K=4$, and the residuals are denoted by $r(\mathbf{s}_i, t_r)=Z(\mathbf{s}_i, t_r)-\sum_{k=1}^Kd_k u_{rk} v_{ik}$.

The sampling locations of the electrostatic potentials are restricted to the high-latitude region of the Northern Hemisphere, while the AXING-need model is constructed for data on a global scale due to the grids on which spherical needlets are placed. There are several ways of handling this problem. First, we can still fit the same model to the data, which are in the high-latitude region, but the needlets outside the region would not have any observations contained in their effective support, and hence the corresponding coefficients could not be effectively estimated from the data \citep[Section 3.1]{Chu-09}. Second, we may simplify the problem as discussed in \citet{heaton-15} through transforming the data domain to a disk in $\mathbb{R}^2$ by the polar projection, but our model needs to be modified accordingly (it is still under the same framework as a random effects model), especially using an appropriate set of basis functions constructed on the disk instead of the sphere. Herein we follow the idea of \citet{Weimer-95} and \citet{ruohoniemi-98}, in which the data are stretched from the high-latitude region to the entire sphere. This is achieved by multiplying the co-latitude $\theta$ with a stretching factor $\alpha$ to obtain a new co-latitude $\theta'=\alpha \theta$, where in our case $\alpha=180/45=4$. When applied to random fields, one drawback of this approach is that the correlation structure may be distorted by the stretching, especially near the low-latitude boundary of the data. However, the magnitude of the electrostatic potentials in the lower latitude region ($\theta > 30^\circ$) is close to zero, and the correlation contours in the higher latitude region ($\theta < 30^\circ$) have elongated shapes that are wider in the east-west direction \citep{Cousins-13}. These findings suggest that the distortion is not as serious as it seems.

We notice that there are still certain large-scale features remaining in the residuals, which cannot be directly modeled by the AXING-need model since it only captures small-scale details at needlet frequency levels higher than or equal to $J_0$. Thus, we consider the following augmented model (the term of observational errors is omitted here)
$$
r(\mathbf{s}_i, t_r)=g(\mathbf{s}_i', t_r) \left( \sum \limits_{l=0}^L \sum \limits_{m=-l}^l a_{lm}(t_r)Y_{lm}(\textbf{s}'_i) + \sum \limits_{j=J_0}^J \sum \limits_{k=1}^{p_j} c_{jk}(t_r)\psi_{jk}(\textbf{s}'_i) \right),
$$
where the first term within the parentheses involving SH functions is included to represent the remaining large-scale features, $\mathbf{s}_i'$ is the point obtained by stretching $\mathbf{s}_i$, and the random coefficients $a_{lm}$ and $c_{jk}$ all depend on the time $t_r$.
To ensure that the SH functions and needlets cover the full frequency range with the minimum overlap, we specify $L=3$ and $J_0=2$ given $B=2$. We also specify $J=4$. This indicates that there are in total three levels of needlets in the model, and the finest details it can capture are at the frequency level $l=31$ in terms of SH functions.
For simplicity, instead of simultaneously estimating the parameters of the SH and needlet terms, we remove a crude estimate of the SH term from the residuals. First, we further assume that the function $g$ does not vary over time, and the residuals are i.i.d. replications over time. These two assumptions are made primarily to obtain a crude estimate of $g$ by the moment estimator, denoted by $\hat{g}(\mathbf{s}'_i)$, based on the replications over time. The residuals are then standardized by dividing by $\hat{g}(\mathbf{s}'_i)$. We filter out the SH term through regressing the standardized residuals on SH functions up to $L=3$. Last, we multiply the residuals after regression with $\hat{g}(\mathbf{s}'_i)$, and treat them as the small-scale component of the electrostatic potentials. Note that a formal statistical analysis on the augmented model is left for future work.

\begin{table}[htp]
\caption{Parameter estimates for the AXING-need ($\nu$=3), Gau-need and Gau-Mat\'ern models applied to the small-scale component of the high-latitude ionospheric electrostatic potential at the first time point. The corresponding standard errors are shown in parentheses. Note that $\eta_0$ is fixed at $0$ for the AXING-need and Gau-need models}
\begin{center}
\begin{tabular}{cccc}
\hline
\hline
Model & AXING-need & Gau-need & Gau-Mat\'ern \\
\hline
$\eta_0$ & 0 (-) & 0 (-) & -2.037 (0.16) \\
$\eta_1$ & 0.949 (0.036) & 0.959 (0.011) & 0.659 (0.10) \\
$\eta_2$ & 1.579 (0.086) & 1.619 (0.012) & 1.372 (0.29) \\
$\eta_3$ & -1.209 (0.046) & -1.190 (0.012) & -1.078 (0.070) \\
$\sigma_2$ & 0.148 (0.023) & 0.252 (0.014) & -\\
$\sigma_3$ & 0.0363 (0.0035) & 0.0665 (0.0032) & - \\
$\sigma_4$ & 4.09e-3 (2.62e-4) & 6.47e-3 (2.11e-04) & -\\
$\kappa$ & - & - & 2.857 (0.14) \\
$1/a$ & - & - & 0.102 (0.0072) \\
$\tau$ & 0.0280 (3.71e-04) & 0.0282 (3.18e-04) & 7.76e-03 (1.16e-04) \\
\hline
\end{tabular}
\end{center}
\label{table:fitted_param}
\end{table}%

\subsection{Model fitting and prediction results}
We fit the AXING-need, Gau-need and Gau-Mat\'ern models to the small-scale component of the electrostatic potential at the first time point (see the top left panel of Figure \ref{fig:real_nonGau}) since we are mainly interested in modeling the small-scale spatial variability. For future work, we may extend our model to spatio-temporal settings under the same framework with additional temporal structures on the coefficients $c_{jk}$'s. Without loss of generality, the Earth is treated as a unit sphere.
The function $g$ is assumed to be a function of co-latitude and the basis functions $b_i(\theta')$'s ($\theta'=4\theta$ due to the stretching) are chosen as natural cubic B-splines with two interior knots and one boundary knot (see Section S.2.2 of the supplementary material \citet{Fan-17-supp} for details). To reduce the computational burden, we randomly select $4000$ locations from the original grid to fit the models. We choose $\nu=3$ among 2.5, 3 and 4 since it yields the best predictive performance (in terms of the MAE, MSPE and CRPS) when predicting at the remaining locations. Moreover, too large $\nu$ ($\nu>4$) makes the model too close to Gaussian, and $\nu$ has to be larger than 2 to ensure the existence of the first two moments. For the AXING-need model, the parameters are estimated by the adaptive MCMC algorithm in Section \ref{sec:MCMC}, in which $\tau_{\bm{\eta}}=10$. The chain was run for 600,000 iterations, with the first 400,000 samples discarded as a burn-in period, and every $200$th sample is collected as the posterior samples. The MCMC diagnostics can be found in Section S.2.3 of the supplementary material \citep{Fan-17-supp}. For the Gaussian models, the parameters are estimated by the maximum likelihood method. Table \ref{table:fitted_param} lists the parameter estimates and the corresponding standard errors for the models. The standard errors are estimated by the posterior sample standard deviations and the parametric bootstrap with $200$ bootstrap samples for the AXING-need model and the Gaussian models, respectively. The parametric bootstrap method was used in spatial settings in \citet{Xu-16, Fan-16}. %It is worth mentioning that the standard error of $\eta_0$ for the Gau-need model, and the standard error of $\eta_i$'s for the Gau-Mat\'ern model are much larger than the counterparts for the AXING-need model.

\begin{table}[htp]
\caption{Predictive performance of the Gau-need and Gau-Mat\'ern models for the small-scale component of the high-latitude ionospheric electrostatic potential at the first time point. All the scores are averaged over 100 cross-validation replications, where the standard deviations are shown in parentheses}
\begin{center}
\resizebox{\textwidth}{!}{%
\begin{tabular}{c|ccc|cccc}
\hline
\hline
Model & MAE (kV) & MSPE (k$\rm V^2$) & CRPS & CP($50\%$) & mLen & CP($90\%$) & mLen \\
\hline
Gau-need & 0.138 (0.024) & 0.146 (0.074) & 0.102 (0.019) & 68.2\% & 0.23 & 92.5\% & 0.57 \\
Gau-Mat\'ern & 0.136 (0.031) & 0.158 (0.092) & 0.113 (0.029) & 43.6\% & 0.08 & 68.8\% & 0.21  \\
\hline
\end{tabular}}
\end{center}
\label{table:kriging_result}
\end{table}%

Apart from model fitting, we further compare the predictive performance of the models. %The simulation results in Section \ref{sec:pred_perf_comp} indicate that the AXING-need and Gau-need models have very similar predictive performance when $\nu=4$. 
To reduce the computational burden, we only compare the Gau-need and Gau-Mat\'ern models since the parameter estimation and spatial prediction for the AXING-need model involves more time-consuming MCMC iterations. 
The scientific implications and benefits of using a non-Gaussian model  compared with Gaussian models will be illustrated in Section \ref{sec:sci_imp}. 
Section S.2.6 of the supplementary material \citep{Fan-17-supp} gives the training and test set separation in a cross-validation procedure, specified according to the spatial coverage of the SuperDARN data. We predict on the test set using the parameter estimates in Table \ref{table:fitted_param}. The scoring rules MAE, MSPE and CPRS are used to assess the prediction accuracy, averaged over all the predicted locations. We repeat the cross-validation procedure 100 times, and calculate the mean and standard deviation (in parentheses) of the scores, which are displayed in Table \ref{table:kriging_result}. There is no significant difference in predictive performance between the two models, with the Gau-need model performing slightly better than the Gau-Mat\'ern model in terms of the MSPE and CRPS. Table \ref{table:kriging_result} also shows the mean lengths of 50\% and 90\% predictive intervals and the corresponding coverage probabilities, averaged over the 100 cross-validation replications. The Gau-need model yields longer predictive intervals, but they have more satisfactory coverage probabilities.

\begin{figure}[htbp] %  figure placement: here, top, bottom, or page
   \centering
   \includegraphics[width=0.8\linewidth]{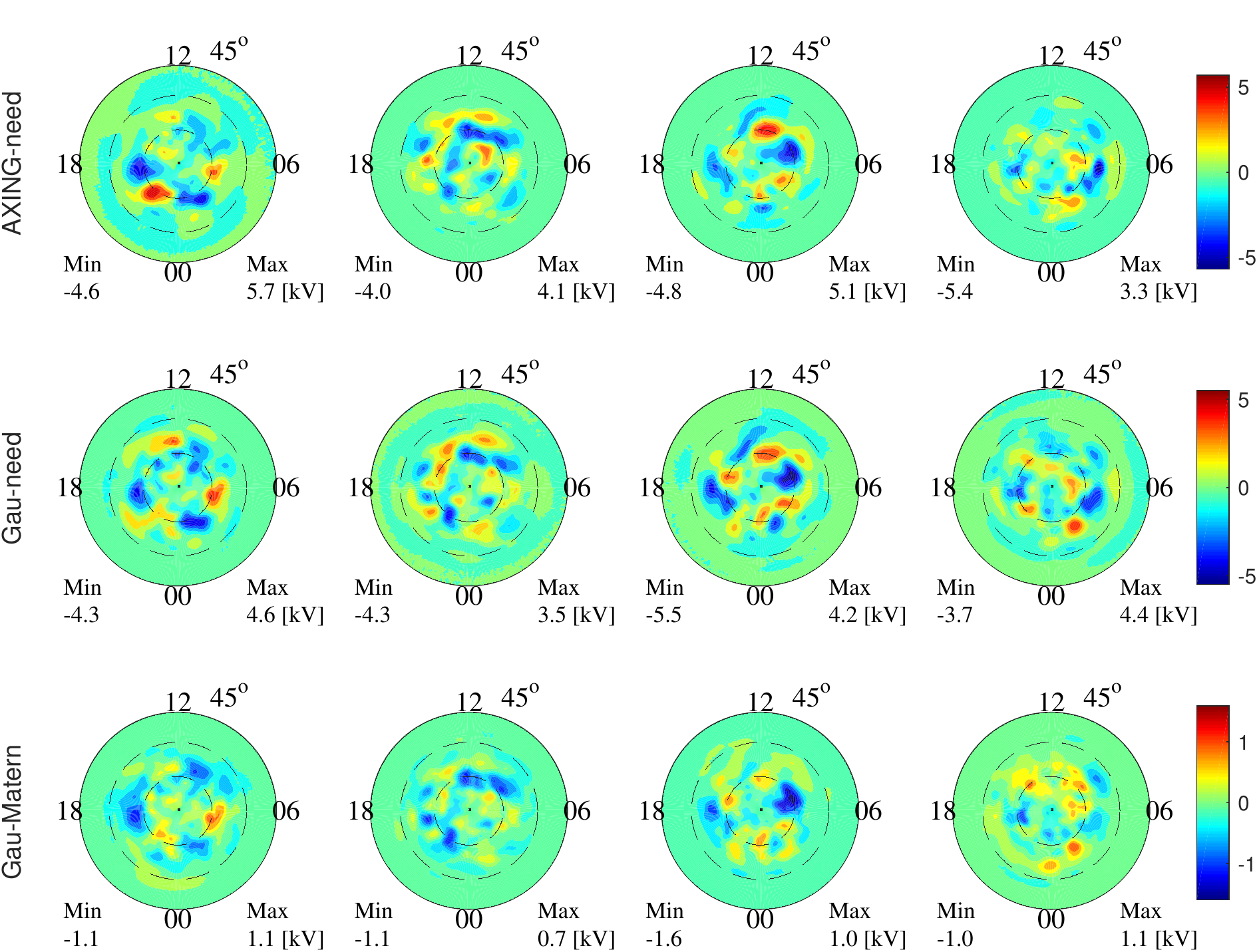} 
   \caption{Comparison of simulations of the AXING-need, Gau-need and Gau-Mat\'ern models.}
   \label{fig:sim_small_scale}
\end{figure}

\subsection{Scientific implications}\label{sec:sci_imp}
As mentioned in Section \ref{sec:intro}, one important scientific goal of modeling the electrostatic potentials is to generate their simulations for estimating high-latitude energy inputs into the upper atmosphere in general circulation models. In Figure \ref{fig:sim_small_scale}, we  compare the simulations of the fitted AXING-need, Gau-need and Gau-Mat\'ern models, for which the parameter estimates are given in Table \ref{table:fitted_param}. These simulations are generated such that their patterns are comparable across the models (see Section S.2.5 of the supplementary material \citet{Fan-17-supp}). We can see that the magnitude of the simulations of the Gau-Mat\'ern model is much smaller than that of the AXING-need and Gau-need models. %while the magnitude of the simulations of the AXING-need model is the largest among the three. 
The most activity (i.e., the red and blue patches in the online version of this figure) of the simulations of all the models occurs near $75^\circ$ latitude, which is consistent with the location of the general auroral zone \citep[Chapter 6.2.1]{Hunsucker-07}. Compared with the Gaussian models, the red and blue patches of the simulations of the AXING-need model have the sharpest and clearest boundaries. This can be attributed to its non-Gaussianity, which makes only a small number of the coefficients $c_{jk}$'s non-zero and the rest of them almost zero. Additional simulations of the fitted AXING-need model are shown in Section S.2.4 of the supplementary material \citep{Fan-17-supp}.

\begin{figure}[htbp] %  figure placement: here, top, bottom, or page
   \centering
   \includegraphics[width=0.8\linewidth]{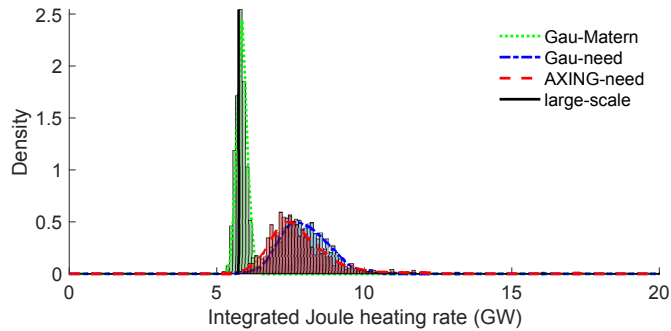} 
   \caption{Histograms of the overall integrated Joule heating rates of $1000$ electrostatic potentials in the high-latitude region of the Northern Hemisphere simulated from the AXING-need, Gan-need and Gau-Mat\'ern models.}
   \label{fig:energy}
\end{figure}
Finally, we quantify the amount of Joule heating produced by the small-scale component of the electrostatic potentials. The estimation of the Joule heating rates (i.e., the amount of Joule heating per second) based on the electrostatic potentials is given in Appendices \ref{sec:electric_field} and \ref{sec:Joule}.
For each fitted model, we compute the overall integrated Joule heating rate of $1000$ simulated electrostatic potentials in the high-latitude region of the Northern Hemisphere, and plot the corresponding histogram in Figure \ref{fig:energy}. The overall integrated Joule heating rates for the AXING-need and Gau-need models are significantly larger than that for the large-scale component only, while the latter is almost the same as the overall integrated Joule heating rates for the Gau-Mat\'ern model. Note that the large-scale component is what has been subtracted from the data in the preprocessing step. Moreover, the overall integrated Joule heating rates for the AXING-need model have a heavier right tail than those for the Gau-need model. Specifically, the 95th and 99th percentiles of the overall integrated Joule heating rates for the AXING-need model are 9.6842 and 11.6769, while those for the Gau-need model are 9.2981 and 9.9659. Consequently, the AXING-need model can produce significantly larger extreme values of the Joule heating rate than the Gau-need model. All of these suggest that the proposed AXING-need model can provide a potentially viable remedy to systematic biases of general circulation models resulting from the underestimation of high-latitude energy inputs.

\section{Discussion}\label{sec:discuss}
We have introduced a new class of multi-resolution spatial models for non-Gaussian  random fields on a sphere. They are constructed in the form of a sparse random effects model using spherical needlets as building blocks. The spatial localization of needlets, together with carefully chosen random coefficients, ensure the model to be non-Gaussian and isotropic. We have shown that the proposed model has an oscillating and quasi-exponentially decaying covariance function. The model has also been expanded to include a spatially varying variance profile.  The special formulation of the model enables us to develop efficient estimation and prediction procedures under a hierarchical Bayesian framework. We have investigated the accuracy of parameter estimation of the proposed model, and compared its predictive performance with that of two Gaussian models by extensive numerical experiments. The effectiveness of the proposed model is also demonstrated through an application of the methodology to the high-latitude ionospheric electrostatic potentials generated from the LFM-MIX model.

The computational speed of each iteration of the adaptive MCMC algorithm is primarily determined by the sampling step for the coefficient vector $\mathbf{c}$, even though we have already partitioned it into subblocks $\mathbf{c}_j$'s to speed up the computation. Recall that $\mathbf{c}_j$ is sampled from $\mathcal{N}(\hat{\bm{\mu}}_j, \widehat{\mathbf{\Sigma}}_j)$. When the number of observations is much smaller than that of needlets, we can reduce the computational burden significantly through applying the Sherman-Morrison-Woodbury formula to $\widehat{\mathbf{\Sigma}}_j$. This coincides with the algorithm proposed in \citet{Bhattacharya-16}.

Many terrestrial processes arising in geophysical and environmental sciences are vector fields tangential to the surface of the Earth. Along the lines of \citet{Fan-16}, extensions to models for tangential vector fields on a sphere are possible through applying spherical differential operators to the proposed model. Then, the tangential vector fields are represented in terms of \textit{vectorial needlets}. The proposed model can also be extended to models for anisotropic random fields, as mentioned in Section \ref{sec:spa_rand_eff}, by allowing $\sigma_{jk}$ and $\nu_{jk}$ to vary spatially, and dynamic spatio-temporal random fields through imposing a time series model on the coefficients $c_{jk}$'s \citep{Cressie-10}.

\appendix

\section{Proof of Theorem 1}\label{sec: proof_prop1}
Since $c_{jk}$'s are independent, we obtain
\begin{eqnarray}
\mbox{Cov}(X(\textbf{s}), X(\textbf{t})) & = & \sum \limits_{j=J_0}^J \sum \limits_{k=1}^{p_j} \mbox{Var}(c_{jk}) \psi_{jk}(\textbf{s}) \psi_{jk}(\textbf{t}) \nonumber \\
& = & \sum \limits_{j=J_0}^J \frac{\nu_j \sigma_j^2}{\nu_j-2} \sum \limits_{k=1}^{p_j} \psi_{jk}(\textbf{s}) \psi_{jk}(\textbf{t}). \nonumber
\end{eqnarray}

        Using (\ref{needlet}), it follows that
	\begin{eqnarray}
	&& \sum_{k=1}^{p_j} \psi_{jk}(\textbf{s}) \psi_{jk}(\textbf{t}) \nonumber\\
	&=&  \sum \limits_{l=\lceil B^{j-1} \rceil}^{\lfloor B^{j+1} \rfloor}
	\sum \limits_{l'=\lceil B^{j-1} \rceil}^{\lfloor B^{j+1} \rfloor} b\left( \frac{l}{B^j} \right) 
	b\left( \frac{l'}{B^j} \right) \left(\frac{2l+1}{4\pi}\right) \left(\frac{2l'+1}{4\pi}\right)
	\sum_{k=1}^{p_j} \lambda_{jk} P_l(\langle \bm{\zeta}_{jk}, \textbf{s}\rangle) P_{l'}(\langle \bm{\zeta}_{jk}, \textbf{t}\rangle) \nonumber \\
	&=&  \sum \limits_{l=\lceil B^{j-1} \rceil}^{\lfloor B^{j+1} \rfloor}
	\sum \limits_{l'=\lceil B^{j-1} \rceil}^{\lfloor B^{j+1} \rfloor} b\left( \frac{l}{B^j} \right) 
	b\left( \frac{l'}{B^j} \right) \left(\frac{2l+1}{4\pi}\right) \left(\frac{2l'+1}{4\pi}\right)
	\int_{\mathbb{S}^2} P_l(\langle \textbf{x}, \textbf{s}\rangle) P_{l'}(\langle \textbf{x}, \textbf{t}\rangle) d\textbf{x}, \nonumber
	\end{eqnarray}
	where the last step is due to the quadrature formula (\ref{quadature}). Now, by the fact that 
	$P_l$ is real-valued and the \textit{Addition Theorem for SH functions},  we have 
	\begin{eqnarray*}
		&&  \left(\frac{2l+1}{4\pi}\right) \left(\frac{2l'+1}{4\pi}\right)
		\int_{\mathbb{S}^2} P_l(\langle \textbf{x}, \textbf{s}\rangle) P_{l'}(\langle \textbf{x}, \textbf{t}\rangle) d\textbf{x}\\
		&=& \int_{\mathbb{S}^2} \sum_{m=-l}^{l}\sum_{m'=-l'}^{l'} \overline{Y}_{lm}(\textbf{x}) Y_{lm}(\textbf{s}) Y_{l'm'}(\textbf{x}) \overline{Y}_{l'm'}(\textbf{t}) d\textbf{x}\\
		&=& \sum_{m=-l}^{l}\sum_{m'=-l'}^{l'} Y_{lm}(\textbf{s})\overline{Y}_{l'm'}(\textbf{t})
		\int_{\mathbb{S}^2} Y_{l'm'}(\textbf{x}) \overline{Y}_{lm}(\textbf{x}) d\textbf{x}\\
		&=& \sum_{m=-l}^{l}\sum_{m'=-l'}^{l'} Y_{lm}(\textbf{s})\overline{Y}_{l'm'}(\textbf{t}) \delta_{ll'}\delta_{mm'}\\
		&=& \delta_{ll'} \sum_{m=-l}^{l} Y_{lm}(\textbf{s})\overline{Y}_{lm}(\textbf{t})\\
		&=& \delta_{ll'}  \left(\frac{2l+1}{4\pi}\right) P_l(\langle \textbf{s}, \textbf{t}\rangle),
	\end{eqnarray*}
	where $\delta_{ab}=1$ if $a=b$, and $0$ otherwise. Note that the third equality follows from the orthonormality of SH functions, and the last one is
	due to another use of the \textit{Addition Theorem}. Substituting this into the above equation, we obtain 
	\[
	\sum_{k=1}^{p_j} \psi_{jk}(\textbf{s}) \psi_{jk}(\textbf{t})
	=  \sum \limits_{l=\lceil B^{j-1} \rceil}^{\lfloor B^{j+1} \rfloor} b^2 \left( \frac{l}{B^j} \right) \left(\frac{2l+1}{4\pi} \right)P_l(\langle \textbf{s}, \textbf{t} \rangle).
	\]

Thus,
$$\mbox{Cov}(X(\textbf{s}), X(\textbf{t}))=\sum \limits_{j=J_0}^J \frac{\nu_j \sigma_j^2}{\nu_j-2}\sum \limits_{l=\lceil B^{j-1} \rceil}^{\lfloor B^{j+1} \rfloor} b^2 \left( \frac{l}{B^j} \right) \left(\frac{2l+1}{4\pi}\right) P_l(\langle \textbf{s}, \textbf{t} \rangle).
$$
By the bound used in the proof of \citet[Lemma 3]{Baldi-09},
$$\left\lvert \sum \limits_{l=\lceil B^{j-1} \rceil}^{\lfloor B^{j+1} \rfloor} b^2 \left( \frac{l}{B^j} \right) \left(\frac{2l+1}{4\pi} \right)P_l(\langle \textbf{s}, \textbf{t} \rangle) \right\rvert \leq \frac{c_M B^{2j}}{[1+B^j \arccos (\langle \textbf{s}, \textbf{t} \rangle)]^M},$$
where $c_M$ is some constant depending on $M$. Note that a direct proof of the above inequality can be derived from \citet[Theorem 3.5]{Narcowich-etal06} or \citet[Theorem 2.6.7]{Dai-13}.

Thus,
$$\lvert \mbox{Cov}(X(\textbf{s}), X(\textbf{t})) \rvert \leq \sum \limits_{j=J_0}^J \frac{\nu_j \sigma_j^2}{\nu_j-2}\frac{c_M B^{2j}}{[1+B^j \arccos (\langle \textbf{s}, \textbf{t} \rangle)]^M}.$$

\section{Deriving Electric Field from Electrostatic Potential}\label{sec:electric_field}
Denote the electric field by $\textbf{E}$ and the electrostatic potential by $\Phi_{\rm E}$. Then
$$\textbf{E}=-\nabla \Phi_{\rm E}\approx-\frac{1}{R}\frac{\partial \Phi_{\rm E}}{\partial \theta}\hat{\bm{\theta}}-\frac{1}{R} \frac{1}{\sin \theta} \frac{\partial \Phi_{\rm E}}{\partial \phi}\hat{\bm{\phi}},$$
where $\nabla$ is the gradient operator defined on $\mathbb{R}^3$, $\hat{\bm{\theta}}$ and $\hat{\bm{\phi}}$ are two of the spherical unit vectors with $\hat{\bm{\theta}}$ and $\hat{\bm{\phi}}$ pointing southward and eastward, respectively, and $R \approx 6.5 \times 10^6 m$ is the radius of the ionosphere. Note that the radial component of the electric field is ignored in this approximation since its magnitude is relatively small compared with that of the tangential component.
Recall that the electrostatic potential in the high-latitude region of the Northern Hemisphere can be expressed as
$$\Phi_{\rm E}(\theta, \phi)=g(\theta')\sum_{j, k}c_{jk}\psi_{jk}(\theta', \phi)=g(4\theta)\sum_{j, k}c_{jk}\psi_{jk}(4\theta, \phi),$$
where $\theta'=4\theta\in [0, \pi]$ due to the stretching of the data to the entire sphere.
Then we have
$$\frac{\partial \Phi_{\rm E}}{\partial \theta}=4 \left(  \frac{\partial g}{\partial \theta'}\bigg{|}_{\theta'=4\theta}\sum_{j, k}c_{jk}\psi_{jk}(4\theta, \phi)+g(4\theta)\sum_{j, k}c_{jk}\frac{\partial \psi_{jk}}{\partial \theta' }\bigg{|}_{\theta'=4\theta}\right),$$
and
$$\frac{1}{\sin \theta}\frac{\partial \Phi_{\rm E}}{\partial \phi}=\frac{\sin \theta'}{\sin \theta}g(4\theta)\sum_{j, k}c_{jk}\frac{1}{\sin \theta'}\frac{\partial \psi_{jk}}{\partial \phi}.$$
%Next, we will compute $\frac{\partial \psi_{jk}}{\partial \theta}$ and $\frac{1}{\sin \theta}\frac{\partial \psi_{jk}}{\partial \phi}$.
Recall that 
$$\psi_{jk}(\theta', \phi)=\sqrt{\lambda_{jk}}\sum_l b\left( \frac{l}{B^j} \right)\frac{2l+1}{4\pi}P_l(x_{jk}\sin \theta' \cos \phi +y_{jk} \sin \theta' \sin \phi + z_{jk}\cos \theta' ).$$
Then
$$\frac{\partial \psi_{jk}}{\partial \theta'}=\sqrt{\lambda_{jk}}(x_{jk}\cos \theta' \cos \phi +y_{jk}\cos \theta' \sin \phi  -z_{jk} \sin \theta' ) \sum_l b\left( \frac{l}{B^j} \right)\frac{2l+1}{4\pi}\frac{d P_l(u)}{d u}\bigg{|}_{u=u'},$$
and
$$\frac{1}{\sin \theta'}\frac{\partial \psi_{jk}}{\partial \phi}=\sqrt{\lambda_{jk}}(-x_{jk} \sin \phi + y_{jk}\cos \phi )\sum_l b\left( \frac{l}{B^j} \right)\frac{2l+1}{4\pi}\frac{d P_l(u)}{d u}\bigg{|}_{u=u'},$$
where $u'=x_{jk}\sin \theta' \cos \phi +y_{jk} \sin \theta' \sin \phi + z_{jk}\cos \theta' $. Note that $dP_l(u)/du$ can be efficiently computed by a recursive formula.

\section{Computing Ionospheric Joule Heating Rate}\label{sec:Joule}
According to \citet{Palmroth-05}, the ionospheric Joule heating rate can be estimated by 
$$\widehat{P}_{\rm JH}(\theta, \phi)=\Sigma_{\rm P}(\theta, \phi)|\mathbf{E}(\theta, \phi)|^2 \approx \Sigma_{\rm P}(\theta, \phi)\left( E_\theta(\theta, \phi)^2+E_\phi(\theta, \phi)^2 \right),$$
where $\Sigma_{\rm P}$ is the height-integrated Pedersen conductivity, and $E_\theta$ and $E_\phi$ are the zonal and meridional components of the electric field given in Appendix \ref{sec:electric_field}, i.e., 
$$E_\theta=-\frac{1}{R}\frac{\partial \Phi_{\rm E}}{\partial \theta},$$
and
$$E_\phi=-\frac{1}{R} \frac{1}{\sin \theta} \frac{\partial \Phi_{\rm E}}{\partial \phi}.$$
The Joule heating rate integrated over a subset $\mathcal{S}$ of the ionosphere is defined as
$$P_{\rm IJH}(\mathcal{S}) = \int_{\mathcal{S}}P_{\rm JH}(\theta, \phi)dS,$$ where 
$dS$ is the area element of the ionosphere. In particular when $\mathcal{S}$ is the high-latitude region of the Northern Hemisphere (co-latitude $\theta \leq \pi/4$),
$$P_{\rm IJH}= \int_{\theta \leq \pi/4}P_{\rm JH}(\theta, \phi)dS.$$

\section*{Acknowledgements}
A part of the work of the first author was done while he was visiting National Center for Atmospheric Research during the summer of 2014, 2015 and 2016. The authors are grateful to Doug Nychka, the editor Nicoleta Serban, the associated editor, and the referees for their valuable comments and suggestions. The data set of the LFM-MIX model output was kindly provided by Mike Wiltberger. The authors also thank Ellen Cousins for her valuable comments and suggestions.

\begin{supplement}
\stitle{Supplementary material for ``A multi-resolution model for non-Gaussian random fields on a sphere with application to ionospheric electrostatic potentials"}
\slink[doi]{COMPLETED BY THE TYPESETTER}
\sdatatype{.pdf}
\sdescription{This supplement provides additional figures and details of the numerical experiments and application.}
\end{supplement}

\bibliography{refs}

\begin{thebibliography}{59}
% BibTex style file: imsart-nameyear.bst, 2013-01-28
% Default style options (sort=1,type=nameyear).
% Used options (sort=1,type=nameyear).

\bibitem[\protect\citeauthoryear{Andrews and Mallows}{1974}]{Andrews-74}
\begin{barticle}[author]
\bauthor{\bsnm{Andrews},~\bfnm{David~F}\binits{D.~F.}} \AND
  \bauthor{\bsnm{Mallows},~\bfnm{Colin~L}\binits{C.~L.}}
(\byear{1974}).
\btitle{Scale mixtures of normal distributions}.
\bjournal{Journal of the Royal Statistical Society. Series B (Methodological)}
\bpages{99--102}.
\end{barticle}
\endbibitem

\bibitem[\protect\citeauthoryear{Andrieu and Thoms}{2008}]{Andrieu-08}
\begin{barticle}[author]
\bauthor{\bsnm{Andrieu},~\bfnm{Christophe}\binits{C.}} \AND
  \bauthor{\bsnm{Thoms},~\bfnm{Johannes}\binits{J.}}
(\byear{2008}).
\btitle{A tutorial on adaptive {MCMC}}.
\bjournal{Statistics and Computing}
\bvolume{18}
\bpages{343-373}.
\bdoi{10.1007/s11222-008-9110-y}
\end{barticle}
\endbibitem

\bibitem[\protect\citeauthoryear{Atkinson and Han}{2012}]{Atkinson-12}
\begin{bbook}[author]
\bauthor{\bsnm{Atkinson},~\bfnm{Kendall}\binits{K.}} \AND
  \bauthor{\bsnm{Han},~\bfnm{Weimin}\binits{W.}}
(\byear{2012}).
\btitle{Spherical harmonics and approximations on the unit sphere: an
  introduction}.
\bpublisher{Springer}, \baddress{Berlin, Heidelberg}.
\end{bbook}
\endbibitem

\bibitem[\protect\citeauthoryear{Baldi et~al.}{2009}]{Baldi-09}
\begin{barticle}[author]
\bauthor{\bsnm{Baldi},~\bfnm{P.}\binits{P.}},
  \bauthor{\bsnm{Kerkyacharian},~\bfnm{G.}\binits{G.}},
  \bauthor{\bsnm{Marinucci},~\bfnm{D.}\binits{D.}} \AND
  \bauthor{\bsnm{Picard},~\bfnm{D.}\binits{D.}}
(\byear{2009}).
\btitle{Asymptotics for spherical needlets}.
\bjournal{The Annals of Statistics}
\bvolume{37}
\bpages{1150--1171}.
\bdoi{10.1214/08-AOS601}
\end{barticle}
\endbibitem

\bibitem[\protect\citeauthoryear{Barndorff-Nielsen}{1979}]{Barndorff-79}
\begin{barticle}[author]
\bauthor{\bsnm{Barndorff-Nielsen},~\bfnm{O.}\binits{O.}}
(\byear{1979}).
\btitle{Models for Non-{Gaussian} Variation, with Applications to Turbulence}.
\bjournal{Proceedings of the Royal Society of London A: Mathematical, Physical
  and Engineering Sciences}
\bvolume{368}
\bpages{501--520}.
\bdoi{10.1098/rspa.1979.0144}
\end{barticle}
\endbibitem

\bibitem[\protect\citeauthoryear{Berg et~al.}{2016}]{Berg-16}
\begin{barticle}[author]
\bauthor{\bsnm{Berg},~\bfnm{Jacob}\binits{J.}},
  \bauthor{\bsnm{Natarajan},~\bfnm{Anand}\binits{A.}},
  \bauthor{\bsnm{Mann},~\bfnm{Jakob}\binits{J.}} \AND
  \bauthor{\bsnm{Patton},~\bfnm{Edward~G}\binits{E.~G.}}
(\byear{2016}).
\btitle{Gaussian vs {non-Gaussian} turbulence: impact on wind turbine loads}.
\bjournal{Wind Energy}.
\end{barticle}
\endbibitem

\bibitem[\protect\citeauthoryear{Bhattacharya, Chakraborty and
  Mallick}{2016}]{Bhattacharya-16}
\begin{barticle}[author]
\bauthor{\bsnm{Bhattacharya},~\bfnm{Anirban}\binits{A.}},
  \bauthor{\bsnm{Chakraborty},~\bfnm{Antik}\binits{A.}} \AND
  \bauthor{\bsnm{Mallick},~\bfnm{Bani~K.}\binits{B.~K.}}
(\byear{2016}).
\btitle{Fast sampling with {Gaussian} scale mixture priors in high-dimensional
  regression}.
\bjournal{Biometrika}
\bvolume{103}
\bpages{985--991}.
\end{barticle}
\endbibitem

\bibitem[\protect\citeauthoryear{Chu, Clyde and Liang}{2009}]{Chu-09}
\begin{barticle}[author]
\bauthor{\bsnm{Chu},~\bfnm{Jen-Hwa}\binits{J.-H.}},
  \bauthor{\bsnm{Clyde},~\bfnm{Merlise~A}\binits{M.~A.}} \AND
  \bauthor{\bsnm{Liang},~\bfnm{Feng}\binits{F.}}
(\byear{2009}).
\btitle{Bayesian function estimation using continuous wavelet dictionaries}.
\bjournal{Statistica Sinica}
\bpages{1419--1438}.
\end{barticle}
\endbibitem

\bibitem[\protect\citeauthoryear{Codrescu, Fuller-Rowell and
  Foster}{1995}]{Codrescu-95}
\begin{barticle}[author]
\bauthor{\bsnm{Codrescu},~\bfnm{M~V}\binits{M.~V.}},
  \bauthor{\bsnm{Fuller-Rowell},~\bfnm{T~J}\binits{T.~J.}} \AND
  \bauthor{\bsnm{Foster},~\bfnm{J~C}\binits{J.~C.}}
(\byear{1995}).
\btitle{On the importance of {E-field} variability for {Joule} heating in the
  high-latitude thermosphere}.
\bjournal{Geophysical Research Letters}
\bvolume{22}
\bpages{2393--2396}.
\end{barticle}
\endbibitem

\bibitem[\protect\citeauthoryear{Codrescu et~al.}{2000}]{Codrescu-00}
\begin{barticle}[author]
\bauthor{\bsnm{Codrescu},~\bfnm{M~V}\binits{M.~V.}},
  \bauthor{\bsnm{Fuller-Rowell},~\bfnm{T~J}\binits{T.~J.}},
  \bauthor{\bsnm{Foster},~\bfnm{J~C}\binits{J.~C.}},
  \bauthor{\bsnm{Holt},~\bfnm{J~M}\binits{J.~M.}} \AND
  \bauthor{\bsnm{Cariglia},~\bfnm{S~J}\binits{S.~J.}}
(\byear{2000}).
\btitle{Electric field variability associated with the {Millstone Hill}
  electric field model}.
\bjournal{Journal of Geophysical Research: Space Physics}
\bvolume{105}
\bpages{5265--5273}.
\end{barticle}
\endbibitem

\bibitem[\protect\citeauthoryear{Cousins, Matsuo and
  Richmond}{2013a}]{Cousins-13-Superdarn}
\begin{barticle}[author]
\bauthor{\bsnm{Cousins},~\bfnm{E~D~P}\binits{E.~D.~P.}},
  \bauthor{\bsnm{Matsuo},~\bfnm{Tomoko}\binits{T.}} \AND
  \bauthor{\bsnm{Richmond},~\bfnm{A~D}\binits{A.~D.}}
(\byear{2013}a).
\btitle{Super{DARN} assimilative mapping}.
\bjournal{Journal of Geophysical Research: Space Physics}
\bvolume{118}
\bpages{7954--7962}.
\end{barticle}
\endbibitem

\bibitem[\protect\citeauthoryear{Cousins, Matsuo and
  Richmond}{2013b}]{Cousins-13}
\begin{barticle}[author]
\bauthor{\bsnm{Cousins},~\bfnm{E.~D.~P.}\binits{E.~D.~P.}},
  \bauthor{\bsnm{Matsuo},~\bfnm{Tomoko}\binits{T.}} \AND
  \bauthor{\bsnm{Richmond},~\bfnm{A.~D.}\binits{A.~D.}}
(\byear{2013}b).
\btitle{Mesoscale and large-scale variability in high-latitude ionospheric
  convection: Dominant modes and spatial/temporal coherence}.
\bjournal{Journal of Geophysical Research: Space Physics}
\bvolume{118}
\bpages{7895--7904}.
\end{barticle}
\endbibitem

\bibitem[\protect\citeauthoryear{Cousins and Shepherd}{2012}]{Cousins-12}
\begin{barticle}[author]
\bauthor{\bsnm{Cousins},~\bfnm{E.~D.~P.}\binits{E.~D.~P.}} \AND
  \bauthor{\bsnm{Shepherd},~\bfnm{S.~G.}\binits{S.~G.}}
(\byear{2012}).
\btitle{Statistical characteristics of small-scale spatial and temporal
  electric field variability in the high-latitude ionosphere}.
\bjournal{Journal of Geophysical Research: Space Physics}
\bvolume{117}.
\end{barticle}
\endbibitem

\bibitem[\protect\citeauthoryear{Cressie}{1993}]{Cressie-93}
\begin{bbook}[author]
\bauthor{\bsnm{Cressie},~\bfnm{Noel}\binits{N.}}
(\byear{1993}).
\btitle{Statistics for Spatial Data}.
\bpublisher{Wiley}, \baddress{New York, NY}.
\end{bbook}
\endbibitem

\bibitem[\protect\citeauthoryear{Cressie and Johannesson}{2008}]{Cressie-08}
\begin{barticle}[author]
\bauthor{\bsnm{Cressie},~\bfnm{Noel}\binits{N.}} \AND
  \bauthor{\bsnm{Johannesson},~\bfnm{Gardar}\binits{G.}}
(\byear{2008}).
\btitle{Fixed rank kriging for very large spatial data sets}.
\bjournal{Journal of the Royal Statistical Society: Series B (Statistical
  Methodology)}
\bvolume{70}
\bpages{209--226}.
\end{barticle}
\endbibitem

\bibitem[\protect\citeauthoryear{Cressie, Shi and Kang}{2010}]{Cressie-10}
\begin{barticle}[author]
\bauthor{\bsnm{Cressie},~\bfnm{Noel}\binits{N.}},
  \bauthor{\bsnm{Shi},~\bfnm{Tao}\binits{T.}} \AND
  \bauthor{\bsnm{Kang},~\bfnm{Emily~L}\binits{E.~L.}}
(\byear{2010}).
\btitle{Fixed rank filtering for spatio-temporal data}.
\bjournal{Journal of Computational and Graphical Statistics}
\bvolume{19}
\bpages{724--745}.
\end{barticle}
\endbibitem

\bibitem[\protect\citeauthoryear{Dai and Xu}{2013}]{Dai-13}
\begin{bbook}[author]
\bauthor{\bsnm{Dai},~\bfnm{Feng}\binits{F.}} \AND
  \bauthor{\bsnm{Xu},~\bfnm{Yuan}\binits{Y.}}
(\byear{2013}).
\btitle{Approximation theory and harmonic analysis on spheres and balls}.
\bpublisher{Springer}, \baddress{New York, NY}.
\end{bbook}
\endbibitem

\bibitem[\protect\citeauthoryear{De~Oliveira, Kedem and Short}{1997}]{De-97}
\begin{barticle}[author]
\bauthor{\bsnm{De~Oliveira},~\bfnm{Victor}\binits{V.}},
  \bauthor{\bsnm{Kedem},~\bfnm{Benjamin}\binits{B.}} \AND
  \bauthor{\bsnm{Short},~\bfnm{David~A}\binits{D.~A.}}
(\byear{1997}).
\btitle{Bayesian prediction of transformed {Gaussian} random fields}.
\bjournal{Journal of the American Statistical Association}
\bvolume{92}
\bpages{1422--1433}.
\end{barticle}
\endbibitem

\bibitem[\protect\citeauthoryear{Fan}{2015}]{Fan-15}
\begin{barticle}[author]
\bauthor{\bsnm{Fan},~\bfnm{Minjie}\binits{M.}}
(\byear{2015}).
\btitle{A Note on Spherical Needlets}.
\bjournal{arXiv preprint arXiv:1508.05406}.
\end{barticle}
\endbibitem

\bibitem[\protect\citeauthoryear{Fan et~al.}{2017a}]{Fan-17-supp}
\begin{barticle}[author]
\bauthor{\bsnm{Fan},~\bfnm{Minjie}\binits{M.}},
  \bauthor{\bsnm{Paul},~\bfnm{Debashis}\binits{D.}},
  \bauthor{\bsnm{Lee},~\bfnm{C.~M.~Thomas}\binits{C.~M.~T.}} \AND
  \bauthor{\bsnm{Matsuo},~\bfnm{Tomoko}\binits{T.}}
(\byear{2017}a).
\btitle{Supplementary material for ``A multi-resolution model for non-Gaussian
  random fields on a sphere with application to ionospheric electrostatic
  potentials"}.
\end{barticle}
\endbibitem

\bibitem[\protect\citeauthoryear{Fan et~al.}{2017b}]{Fan-16}
\begin{barticle}[author]
\bauthor{\bsnm{Fan},~\bfnm{Minjie}\binits{M.}},
  \bauthor{\bsnm{Paul},~\bfnm{Debashis}\binits{D.}},
  \bauthor{\bsnm{Lee},~\bfnm{C.~M.~Thomas}\binits{C.~M.~T.}} \AND
  \bauthor{\bsnm{Matsuo},~\bfnm{Tomoko}\binits{T.}}
(\byear{2017}b).
\btitle{Modeling Tangential Vector Fields on a Sphere}.
\bjournal{Journal of the American Statistical Association}
\bvolume{just-accepted}.
\end{barticle}
\endbibitem

\bibitem[\protect\citeauthoryear{Gelman, Roberts and Gilks}{1996}]{Gelman-96}
\begin{barticle}[author]
\bauthor{\bsnm{Gelman},~\bfnm{Andrew}\binits{A.}},
  \bauthor{\bsnm{Roberts},~\bfnm{Gareth~O}\binits{G.~O.}} \AND
  \bauthor{\bsnm{Gilks},~\bfnm{Walter~R}\binits{W.~R.}}
(\byear{1996}).
\btitle{Efficient {Metropolis} jumping rules}.
\bjournal{Bayesian statistics}
\bvolume{5}
\bpages{599-608}.
\end{barticle}
\endbibitem

\bibitem[\protect\citeauthoryear{Genton and Kleiber}{2015}]{Genton-15}
\begin{barticle}[author]
\bauthor{\bsnm{Genton},~\bfnm{Marc~G.}\binits{M.~G.}} \AND
  \bauthor{\bsnm{Kleiber},~\bfnm{William}\binits{W.}}
(\byear{2015}).
\btitle{Cross-Covariance Functions for Multivariate Geostatistics}.
\bjournal{Statistical Science}
\bvolume{30}
\bpages{147--163}.
\end{barticle}
\endbibitem

\bibitem[\protect\citeauthoryear{Gneiting}{2013}]{Gneiting-13}
\begin{barticle}[author]
\bauthor{\bsnm{Gneiting},~\bfnm{Tilmann}\binits{T.}}
(\byear{2013}).
\btitle{Strictly and non-strictly positive definite functions on spheres}.
\bjournal{Bernoulli}
\bvolume{19}
\bpages{1327--1349}.
\end{barticle}
\endbibitem

\bibitem[\protect\citeauthoryear{Gneiting, Balabdaoui and
  Raftery}{2007}]{Gneiting-07-Sharpness}
\begin{barticle}[author]
\bauthor{\bsnm{Gneiting},~\bfnm{Tilmann}\binits{T.}},
  \bauthor{\bsnm{Balabdaoui},~\bfnm{Fadoua}\binits{F.}} \AND
  \bauthor{\bsnm{Raftery},~\bfnm{Adrian~E}\binits{A.~E.}}
(\byear{2007}).
\btitle{Probabilistic forecasts, calibration and sharpness}.
\bjournal{Journal of the Royal Statistical Society: Series B (Statistical
  Methodology)}
\bvolume{69}
\bpages{243--268}.
\end{barticle}
\endbibitem

\bibitem[\protect\citeauthoryear{Gneiting and Raftery}{2007}]{Gneiting-07}
\begin{barticle}[author]
\bauthor{\bsnm{Gneiting},~\bfnm{Tilmann}\binits{T.}} \AND
  \bauthor{\bsnm{Raftery},~\bfnm{Adrian~E}\binits{A.~E.}}
(\byear{2007}).
\btitle{Strictly proper scoring rules, prediction, and estimation}.
\bjournal{Journal of the American Statistical Association}
\bvolume{102}
\bpages{359--378}.
\end{barticle}
\endbibitem

\bibitem[\protect\citeauthoryear{Gneiting and Ranjan}{2011}]{gneiting-11}
\begin{barticle}[author]
\bauthor{\bsnm{Gneiting},~\bfnm{Tilmann}\binits{T.}} \AND
  \bauthor{\bsnm{Ranjan},~\bfnm{Roopesh}\binits{R.}}
(\byear{2011}).
\btitle{Comparing density forecasts using threshold-and quantile-weighted
  scoring rules}.
\bjournal{Journal of Business \& Economic Statistics}
\bvolume{29}
\bpages{411--422}.
\end{barticle}
\endbibitem

\bibitem[\protect\citeauthoryear{G\'orski et~al.}{2005}]{Gorski-05}
\begin{barticle}[author]
\bauthor{\bsnm{G\'orski},~\bfnm{K.~M.}\binits{K.~M.}},
  \bauthor{\bsnm{Hivon},~\bfnm{E.}\binits{E.}},
  \bauthor{\bsnm{Banday},~\bfnm{A.~J.}\binits{A.~J.}},
  \bauthor{\bsnm{Wandelt},~\bfnm{B.~D.}\binits{B.~D.}},
  \bauthor{\bsnm{Hansen},~\bfnm{F.~K.}\binits{F.~K.}},
  \bauthor{\bsnm{Reinecke},~\bfnm{M.}\binits{M.}} \AND
  \bauthor{\bsnm{Bartelmann},~\bfnm{M.}\binits{M.}}
(\byear{2005}).
\btitle{{HEALPix}: A Framework for High-Resolution Discretization and Fast
  Analysis of Data Distributed on the Sphere}.
\bjournal{The Astrophysical Journal}
\bvolume{622}
\bpages{759}.
\end{barticle}
\endbibitem

\bibitem[\protect\citeauthoryear{Guinness and Fuentes}{2016}]{Guinness-16}
\begin{barticle}[author]
\bauthor{\bsnm{Guinness},~\bfnm{Joseph}\binits{J.}} \AND
  \bauthor{\bsnm{Fuentes},~\bfnm{Montserrat}\binits{M.}}
(\byear{2016}).
\btitle{Isotropic covariance functions on spheres: Some properties and modeling
  considerations}.
\bjournal{Journal of Multivariate Analysis}
\bvolume{143}
\bpages{143--152}.
\end{barticle}
\endbibitem

\bibitem[\protect\citeauthoryear{Heaton et~al.}{2015}]{heaton-15}
\begin{barticle}[author]
\bauthor{\bsnm{Heaton},~\bfnm{Matthew~J}\binits{M.~J.}},
  \bauthor{\bsnm{Kleiber},~\bfnm{William}\binits{W.}},
  \bauthor{\bsnm{Sain},~\bfnm{Stephan~R}\binits{S.~R.}} \AND
  \bauthor{\bsnm{Wiltberger},~\bfnm{Michael}\binits{M.}}
(\byear{2015}).
\btitle{Emulating and calibrating the multiple-fidelity {Lyon-Fedder-Mobarry}
  magnetosphere-ionosphere coupled computer model}.
\bjournal{Journal of the Royal Statistical Society: Series C (Applied
  Statistics)}
\bvolume{64}
\bpages{93--113}.
\end{barticle}
\endbibitem

\bibitem[\protect\citeauthoryear{Hunsucker and Hargreaves}{2007}]{Hunsucker-07}
\begin{bbook}[author]
\bauthor{\bsnm{Hunsucker},~\bfnm{Robert~D}\binits{R.~D.}} \AND
  \bauthor{\bsnm{Hargreaves},~\bfnm{John~Keith}\binits{J.~K.}}
(\byear{2007}).
\btitle{The High-Latitude Ionosphere and its Effects on Radio Propagation}.
\bpublisher{Cambridge University Press}, \baddress{Cambridge}.
\end{bbook}
\endbibitem

\bibitem[\protect\citeauthoryear{Jones}{1963}]{Jones-63}
\begin{barticle}[author]
\bauthor{\bsnm{Jones},~\bfnm{Richard~H.}\binits{R.~H.}}
(\byear{1963}).
\btitle{Stochastic Processes on a Sphere}.
\bjournal{The Annals of Mathematical Statistics}
\bvolume{34}
\bpages{213--218}.
\bdoi{10.1214/aoms/1177704257}
\end{barticle}
\endbibitem

\bibitem[\protect\citeauthoryear{Jun and Stein}{2008}]{Jun-08}
\begin{barticle}[author]
\bauthor{\bsnm{Jun},~\bfnm{Mikyoung}\binits{M.}} \AND
  \bauthor{\bsnm{Stein},~\bfnm{Michael~L}\binits{M.~L.}}
(\byear{2008}).
\btitle{Nonstationary covariance models for global data}.
\bjournal{The Annals of Applied Statistics}
\bvolume{2}
\bpages{1271--1289}.
\end{barticle}
\endbibitem

\bibitem[\protect\citeauthoryear{Kleiber et~al.}{2013}]{kleiber-13}
\begin{barticle}[author]
\bauthor{\bsnm{Kleiber},~\bfnm{William}\binits{W.}},
  \bauthor{\bsnm{Sain},~\bfnm{Stephan~R}\binits{S.~R.}},
  \bauthor{\bsnm{Heaton},~\bfnm{Matthew~J}\binits{M.~J.}},
  \bauthor{\bsnm{Wiltberger},~\bfnm{Michael}\binits{M.}},
  \bauthor{\bsnm{Reese},~\bfnm{C~Shane}\binits{C.~S.}},
  \bauthor{\bsnm{Bingham},~\bfnm{Derek}\binits{D.}} \betal{et~al.}
(\byear{2013}).
\btitle{Parameter tuning for a multi-fidelity dynamical model of the
  magnetosphere}.
\bjournal{The Annals of Applied Statistics}
\bvolume{7}
\bpages{1286--1310}.
\end{barticle}
\endbibitem

\bibitem[\protect\citeauthoryear{Kleiber et~al.}{2016}]{Kleiber-16}
\begin{barticle}[author]
\bauthor{\bsnm{Kleiber},~\bfnm{W}\binits{W.}},
  \bauthor{\bsnm{Hendershott},~\bfnm{B}\binits{B.}},
  \bauthor{\bsnm{Sain},~\bfnm{S~R}\binits{S.~R.}} \AND
  \bauthor{\bsnm{Wiltberger},~\bfnm{M}\binits{M.}}
(\byear{2016}).
\btitle{Feature-based validation of the {Lyon-Fedder-Mobarry}
  magnetohydrodynamical model}.
\bjournal{Journal of Geophysical Research: Space Physics}.
\end{barticle}
\endbibitem

\bibitem[\protect\citeauthoryear{Lindgren, Rue and
  Lindstr{\"o}m}{2011}]{Lindgren-11}
\begin{barticle}[author]
\bauthor{\bsnm{Lindgren},~\bfnm{Finn}\binits{F.}},
  \bauthor{\bsnm{Rue},~\bfnm{H{\aa}vard}\binits{H.}} \AND
  \bauthor{\bsnm{Lindstr{\"o}m},~\bfnm{Johan}\binits{J.}}
(\byear{2011}).
\btitle{An explicit link between {Gaussian} fields and {Gaussian Markov} random
  fields: the stochastic partial differential equation approach}.
\bjournal{Journal of the Royal Statistical Society: Series B (Statistical
  Methodology)}
\bvolume{73}
\bpages{423--498}.
\end{barticle}
\endbibitem

\bibitem[\protect\citeauthoryear{Lyon, Fedder and Mobarry}{2004}]{Lyon-04}
\begin{barticle}[author]
\bauthor{\bsnm{Lyon},~\bfnm{J.~G.}\binits{J.~G.}},
  \bauthor{\bsnm{Fedder},~\bfnm{J.~A.}\binits{J.~A.}} \AND
  \bauthor{\bsnm{Mobarry},~\bfnm{C.~M.}\binits{C.~M.}}
(\byear{2004}).
\btitle{The {Lyon}--{Fedder}--{Mobarry} {(LFM)} global {MHD} magnetospheric
  simulation code}.
\bjournal{Journal of Atmospheric and Solar-Terrestrial Physics}
\bvolume{66}
\bpages{1333--1350}.
\end{barticle}
\endbibitem

\bibitem[\protect\citeauthoryear{Marinucci and Peccati}{2011}]{Marinucci2011}
\begin{bbook}[author]
\bauthor{\bsnm{Marinucci},~\bfnm{Domenico}\binits{D.}} \AND
  \bauthor{\bsnm{Peccati},~\bfnm{Giovanni}\binits{G.}}
(\byear{2011}).
\btitle{Random Fields on the Sphere: Representation, Limit Theorems and
  Cosmological Applications}.
\bpublisher{Cambridge University Press}, \baddress{Cambridge}.
\end{bbook}
\endbibitem

\bibitem[\protect\citeauthoryear{Matsuo, Richmond and Nychka}{2002}]{Matsuo-02}
\begin{barticle}[author]
\bauthor{\bsnm{Matsuo},~\bfnm{Tomoko}\binits{T.}},
  \bauthor{\bsnm{Richmond},~\bfnm{Arthur~D}\binits{A.~D.}} \AND
  \bauthor{\bsnm{Nychka},~\bfnm{Douglas~W}\binits{D.~W.}}
(\byear{2002}).
\btitle{Modes of high-latitude electric field variability derived from {DE-2}
  measurements: {Empirical Orthogonal Function} ({EOF}) analysis}.
\bjournal{Geophysical Research Letters}
\bvolume{29}
\bpages{11-1--11-4}.
\end{barticle}
\endbibitem

\bibitem[\protect\citeauthoryear{Matsuo, Richmond and Hensel}{2003}]{Matsuo-03}
\begin{barticle}[author]
\bauthor{\bsnm{Matsuo},~\bfnm{Tomoko}\binits{T.}},
  \bauthor{\bsnm{Richmond},~\bfnm{Arthur~D}\binits{A.~D.}} \AND
  \bauthor{\bsnm{Hensel},~\bfnm{Kristine}\binits{K.}}
(\byear{2003}).
\btitle{High-latitude ionospheric electric field variability and electric
  potential derived from {DE-2} plasma drift measurements: {Dependence} on
  {IMF} and dipole tilt}.
\bjournal{Journal of Geophysical Research: Space Physics}
\bvolume{108}.
\end{barticle}
\endbibitem

\bibitem[\protect\citeauthoryear{Matsuo and Richmond}{2008}]{Matsuo-08}
\begin{barticle}[author]
\bauthor{\bsnm{Matsuo},~\bfnm{Tomoko}\binits{T.}} \AND
  \bauthor{\bsnm{Richmond},~\bfnm{Arthur~D}\binits{A.~D.}}
(\byear{2008}).
\btitle{Effects of high-latitude ionospheric electric field variability on
  global thermospheric {Joule} heating and mechanical energy transfer rate}.
\bjournal{Journal of Geophysical Research: Space Physics}
\bvolume{113}.
\end{barticle}
\endbibitem

\bibitem[\protect\citeauthoryear{Narcowich, Petrushev and
  Ward}{2006}]{Narcowich-etal06}
\begin{barticle}[author]
\bauthor{\bsnm{Narcowich},~\bfnm{Francis~J}\binits{F.~J.}},
  \bauthor{\bsnm{Petrushev},~\bfnm{P}\binits{P.}} \AND
  \bauthor{\bsnm{Ward},~\bfnm{Joseph~D}\binits{J.~D.}}
(\byear{2006}).
\btitle{Localized tight frames on spheres}.
\bjournal{SIAM Journal on Mathematical Analysis}
\bvolume{38}
\bpages{574--594}.
\end{barticle}
\endbibitem

\bibitem[\protect\citeauthoryear{Palacios and Steel}{2006}]{Palacios-06}
\begin{barticle}[author]
\bauthor{\bsnm{Palacios},~\bfnm{M.~Blanca}\binits{M.~B.}} \AND
  \bauthor{\bsnm{Steel},~\bfnm{Mark F.~J}\binits{M.~F.~J.}}
(\byear{2006}).
\btitle{Non-{Gaussian} {Bayesian} Geostatistical Modeling}.
\bjournal{Journal of the American Statistical Association}
\bvolume{101}
\bpages{604-618}.
\end{barticle}
\endbibitem

\bibitem[\protect\citeauthoryear{Palmroth et~al.}{2005}]{Palmroth-05}
\begin{binproceedings}[author]
\bauthor{\bsnm{Palmroth},~\bfnm{M}\binits{M.}},
  \bauthor{\bsnm{Janhunen},~\bfnm{P}\binits{P.}},
  \bauthor{\bsnm{Pulkkinen},~\bfnm{T~I}\binits{T.~I.}},
  \bauthor{\bsnm{Aksnes},~\bfnm{A}\binits{A.}},
  \bauthor{\bsnm{Lu},~\bfnm{G}\binits{G.}},
  \bauthor{\bsnm{{\O}stgaard},~\bfnm{N}\binits{N.}},
  \bauthor{\bsnm{Watermann},~\bfnm{J}\binits{J.}},
  \bauthor{\bsnm{Reeves},~\bfnm{G~D}\binits{G.~D.}} \AND
  \bauthor{\bsnm{Germany},~\bfnm{G~A}\binits{G.~A.}}
(\byear{2005}).
\btitle{Assessment of ionospheric {Joule} heating by {GUMICS-4} {MHD}
  simulation, {AMIE}, and satellite-based statistics: towards a synthesis}.
In \bbooktitle{Annales Geophysicae}
\bvolume{23}
\bpages{2051--2068}.
\end{binproceedings}
\endbibitem

\bibitem[\protect\citeauthoryear{Perron and Sura}{2013}]{Perron-13}
\begin{barticle}[author]
\bauthor{\bsnm{Perron},~\bfnm{Maxime}\binits{M.}} \AND
  \bauthor{\bsnm{Sura},~\bfnm{Philip}\binits{P.}}
(\byear{2013}).
\btitle{Climatology of {non-Gaussian} atmospheric statistics}.
\bjournal{Journal of Climate}
\bvolume{26}
\bpages{1063--1083}.
\end{barticle}
\endbibitem

\bibitem[\protect\citeauthoryear{R{\o}islien and Omre}{2006}]{Roislien-06}
\begin{barticle}[author]
\bauthor{\bsnm{R{\o}islien},~\bfnm{Jo}\binits{J.}} \AND
  \bauthor{\bsnm{Omre},~\bfnm{Henning}\binits{H.}}
(\byear{2006}).
\btitle{T-distributed Random Fields: A Parametric Model for Heavy-tailed
  Well-log Data}.
\bjournal{Mathematical Geology}
\bvolume{38}
\bpages{821--849}.
\end{barticle}
\endbibitem

\bibitem[\protect\citeauthoryear{Ruohoniemi and Baker}{1998}]{ruohoniemi-98}
\begin{barticle}[author]
\bauthor{\bsnm{Ruohoniemi},~\bfnm{J~M}\binits{J.~M.}} \AND
  \bauthor{\bsnm{Baker},~\bfnm{K~B}\binits{K.~B.}}
(\byear{1998}).
\btitle{Large-scale imaging of high-latitude convection with {Super Dual
  Auroral Radar Network HF} radar observations}.
\bjournal{Journal of Geophysical Research: Space Physics}
\bvolume{103}
\bpages{20797--20811}.
\end{barticle}
\endbibitem

\bibitem[\protect\citeauthoryear{Schoenberg}{1942}]{Schoenberg-42}
\begin{barticle}[author]
\bauthor{\bsnm{Schoenberg},~\bfnm{I.~J.}\binits{I.~J.}}
(\byear{1942}).
\btitle{Positive definite functions on spheres}.
\bjournal{Duke Mathematical Journal}
\bvolume{9}
\bpages{96--108}.
\bdoi{10.1215/S0012-7094-42-00908-6}
\end{barticle}
\endbibitem

\bibitem[\protect\citeauthoryear{Stein}{1988}]{Stein-88}
\begin{barticle}[author]
\bauthor{\bsnm{Stein},~\bfnm{Michael~L}\binits{M.~L.}}
(\byear{1988}).
\btitle{Asymptotically efficient prediction of a random field with a
  misspecified covariance function}.
\bjournal{The Annals of Statistics}
\bvolume{16}
\bpages{55--63}.
\end{barticle}
\endbibitem

\bibitem[\protect\citeauthoryear{Stein}{2007}]{Stein-07}
\begin{barticle}[author]
\bauthor{\bsnm{Stein},~\bfnm{Michael~L.}\binits{M.~L.}}
(\byear{2007}).
\btitle{Spatial variation of total column ozone on a global scale}.
\bjournal{The Annals of Applied Statistics}
\bvolume{1}
\bpages{191--210}.
\bdoi{10.1214/07-AOAS106}
\end{barticle}
\endbibitem

\bibitem[\protect\citeauthoryear{Terdik et~al.}{2015}]{Terdik-15}
\begin{barticle}[author]
\bauthor{\bsnm{Terdik},~\bfnm{Gy{\"o}rgy}\binits{G.}} \betal{et~al.}
(\byear{2015}).
\btitle{Angular spectra for {non-Gaussian} isotropic fields}.
\bjournal{Brazilian Journal of Probability and Statistics}
\bvolume{29}
\bpages{833--865}.
\end{barticle}
\endbibitem

\bibitem[\protect\citeauthoryear{Wallin and Bolin}{2015}]{Wallin-15}
\begin{barticle}[author]
\bauthor{\bsnm{Wallin},~\bfnm{Jonas}\binits{J.}} \AND
  \bauthor{\bsnm{Bolin},~\bfnm{David}\binits{D.}}
(\byear{2015}).
\btitle{Geostatistical Modelling Using Non-{Gaussian} {Mat\'ern} Fields}.
\bjournal{Scandinavian Journal of Statistics}
\bvolume{42}
\bpages{872--890}.
\end{barticle}
\endbibitem

\bibitem[\protect\citeauthoryear{Weimer}{1995}]{Weimer-95}
\begin{barticle}[author]
\bauthor{\bsnm{Weimer},~\bfnm{D~R}\binits{D.~R.}}
(\byear{1995}).
\btitle{Models of high-latitude electric potentials derived with a least error
  fit of spherical harmonic coefficients}.
\bjournal{Journal of Geophysical Research: Space Physics}
\bvolume{100}
\bpages{19595--19607}.
\end{barticle}
\endbibitem

\bibitem[\protect\citeauthoryear{West}{1987}]{West-87}
\begin{barticle}[author]
\bauthor{\bsnm{West},~\bfnm{Mike}\binits{M.}}
(\byear{1987}).
\btitle{On scale mixtures of normal distributions}.
\bjournal{Biometrika}
\bvolume{74}
\bpages{646--648}.
\end{barticle}
\endbibitem

\bibitem[\protect\citeauthoryear{Wiltberger et~al.}{2016}]{Wiltberger-16}
\begin{barticle}[author]
\bauthor{\bsnm{Wiltberger},~\bfnm{M}\binits{M.}},
  \bauthor{\bsnm{Rigler},~\bfnm{E~J}\binits{E.~J.}},
  \bauthor{\bsnm{Merkin},~\bfnm{V}\binits{V.}} \AND
  \bauthor{\bsnm{Lyon},~\bfnm{J~G}\binits{J.~G.}}
(\byear{2016}).
\btitle{Structure of High Latitude Currents in Magnetosphere-Ionosphere
  Models}.
\bjournal{Space Science Reviews}
\bpages{1--24}.
\end{barticle}
\endbibitem

\bibitem[\protect\citeauthoryear{Wolfe, Godsill and Ng}{2004}]{Wolfe-04}
\begin{barticle}[author]
\bauthor{\bsnm{Wolfe},~\bfnm{Patrick~J.}\binits{P.~J.}},
  \bauthor{\bsnm{Godsill},~\bfnm{Simon~J.}\binits{S.~J.}} \AND
  \bauthor{\bsnm{Ng},~\bfnm{Wee~Jing}\binits{W.~J.}}
(\byear{2004}).
\btitle{Bayesian variable selection and regularization for time--frequency
  surface estimation}.
\bjournal{Journal of the Royal Statistical Society: Series B (Statistical
  Methodology)}
\bvolume{66}
\bpages{575--589}.
\bdoi{10.1111/j.1467-9868.2004.02052.x}
\end{barticle}
\endbibitem

\bibitem[\protect\citeauthoryear{Womersley}{2015}]{Womersley-15}
\begin{bmisc}[author]
\bauthor{\bsnm{Womersley},~\bfnm{R.~S.}\binits{R.~S.}}
(\byear{2015}).
\btitle{Efficient Spherical Designs with Good Geometric Properties}.
\bhowpublished{\url{http://web.maths.unsw.edu.au/~rsw/Sphere/EffSphDes/}}.
\end{bmisc}
\endbibitem

\bibitem[\protect\citeauthoryear{Xu and Genton}{2017}]{Xu-16}
\begin{barticle}[author]
\bauthor{\bsnm{Xu},~\bfnm{Ganggang}\binits{G.}} \AND
  \bauthor{\bsnm{Genton},~\bfnm{Marc~G}\binits{M.~G.}}
(\byear{2017}).
\btitle{Tukey g-and-h Random Fields}.
\bjournal{Journal of the American Statistical Association}
\bpages{1--14}.
\end{barticle}
\endbibitem

\bibitem[\protect\citeauthoryear{Zhang et~al.}{2011}]{Zhang-11}
\begin{barticle}[author]
\bauthor{\bsnm{Zhang},~\bfnm{B}\binits{B.}},
  \bauthor{\bsnm{Lotko},~\bfnm{W}\binits{W.}},
  \bauthor{\bsnm{Wiltberger},~\bfnm{M~J}\binits{M.~J.}},
  \bauthor{\bsnm{Brambles},~\bfnm{O~J}\binits{O.~J.}} \AND
  \bauthor{\bsnm{Damiano},~\bfnm{P~A}\binits{P.~A.}}
(\byear{2011}).
\btitle{A statistical study of magnetosphere--ionosphere coupling in the
  {Lyon}--{Fedder}--{Mobarry} global {MHD} model}.
\bjournal{Journal of Atmospheric and Solar-Terrestrial Physics}
\bvolume{73}
\bpages{686--702}.
\end{barticle}
\endbibitem

\end{thebibliography}
\bibliographystyle{imsart-nameyear}

\end{document}